\documentclass[apj,twocolumn,twocolappendix,numberedappendix,linenumbers]{openjournal}
\usepackage{lastpage}
\usepackage{natbib} 
\usepackage[dvipsnames]{xcolor}

\usepackage{amsmath}	
\usepackage[title]{appendix}
\usepackage{hyperref}	
\hypersetup{
    colorlinks=true,
    linkcolor=blue,
    citecolor=blue,
    filecolor=blue,
    urlcolor=blue
}
\usepackage[caption=false]{subfig}
\usepackage{soul}        
\usepackage{xspace}
\usepackage{booktabs}
\usepackage{savesym}
\savesymbol{tablenum}
\usepackage{siunitx}
\restoresymbol{SIX}{tablenum}

\usepackage{textgreek}
\usepackage[utf8]{inputenc}
\usepackage[english]{babel}
\usepackage{color,colortbl}
\usepackage{tensind}
\tensordelimiter{?}
\DeclareGraphicsExtensions{.bmp,.png,.jpg,.pdf}
\usepackage{verbatim}
\usepackage[normalem]{ulem}
\usepackage{orcidlink}
\usepackage{graphicx}	

\def\hide#1{}
\newcommand{\eg}{e.g.,}
\newcommand{\ie}{i.e.,}

\newcommand{\mrm}[1]{\mathrm{#1}}
\newcommand{\HI}{\mbox{H\:{\sc i}}\:}

\newcommand{\HeI}{\mbox{He\:{\sc i}}\:}
\newcommand{\HeII}{He\:{\sc ii}\:}

\defcitealias{garcia_star_2023}{G23}

\graphicspath{ {./figs/} } 

\begin{document}





\title{Seeding Cores: A Pathway for Nuclear Star Clusters from Bound Star Clusters in the First Billion Years  \vspace{-15mm}}

\author{Fred Angelo Batan Garcia\orcidlink{0000-0002-4545-2700}$^{1,\star}$ }
\author{Massimo Ricotti\orcidlink{0000-0003-4223-7324}$^{2, \dagger}$}
\author{Kazuyuki Sugimura\orcidlink{0000-0001-7842-5488}$^{3}$}

\thanks{E-mails:$^\star$g.fred@columbia.edu (FABG),
$^\dagger$ricotti@umd.edu (MR)}

\affiliation{
$^{1}$ Department of Astronomy, Columbia University, 538 W 120th Street, New York, NY 10027, USA\\
$^{2}$ Department of Astronomy, University of Maryland, College Park, MD 20742, USA\\
$^{3}$Faculty of Science, Hokkaido University, Sapporo, Hokkaido 060-0810, Japan
}

\shorttitle{Bound and nuclear star cluster formation at high-$z$}
\shortauthors{F.A.B. Garcia, M. Ricotti, and K. Sugimura}

\begin{abstract}
We model the formation of star clusters in a dwarf galaxy progenitor during the first 700 Myr of cosmic history using a cosmological radiation-hydrodynamic simulation with a sub-grid star formation efficiency (SFE) model calibrated from AU-scale radiation-MHD simulations of molecular clouds with varying mass, density, and metallicity. In comparison to a constant SFE model, our model yields more bursty star formation, a more abundant massive star cluster population, and overall a higher stellar mass. Clouds reach SFEs up to 80\%, forming bound star clusters (densities $\sim10^{2-4} ~{\rm M_\odot\:pc^{-2}}$, radii $\lesssim 3~{\rm pc}$) resembling those observed by the James Webb Space Telescope (JWST) in strongly lensed galaxies. Star clusters follow a flat power-law mass function ${\rm d}N/{\rm d}\log M \propto M^\Gamma$ with slope $\Gamma \sim -0.4$. The most massive star clusters ($10^{4-5} ~{\rm M_\odot}$) grow through mergers and have metallicity spreads of $0.05 - 0.1$ dex that roughly scale with mass. The second burst of star formation produce loosely bound star clusters with higher metallicities: $-1.95 < \log(Z/{\rm Z_\odot}) < -1.50$ at lower SFEs ($2 - 20\%$). At $z \sim 8.7$, a nuclear star cluster (NSC) is seeded, growing $83\%$ of its mass ($ 2.4 \times 10^5 ~{\rm M_\odot}$, $20\%$ of the galaxy's stellar mass) through mergers with pre-existing clusters and the rest through in-situ star formation. The early formation of NSCs has interesting implications for seeding supermassive black holes and the population of \textit{little red dots} recently discovered by JWST at $z \gtrsim 5$.
\end{abstract}


\maketitle
\section{Introduction}
Stars in our Galaxy form in star clusters which can either be gravitationally bound (\eg~globular clusters) or unbound (\eg~open clusters). Star formation likely occurs in a similar way at high redshifts during the formation of Population III (Pop III) stars in primordial halos ($z \gtrsim 12$) \citep{bromm_first_2011} and subsequent Population II (Pop II) stars in early galaxies ($z \gtrsim 6$) \citep{adamo_star_2020, klessen_first_2023}. The recent launch of the JWST \citep{Gardner2023PASP..135f8001G} has made –– and continues to make –– progress in studying the high-$z$ frontier \citep[\eg][]{Finkelstein2023, Kokorev2024}; particularly, in the detection of compact star clusters (tens of parsecs to parsec in size) at $z\sim6-10$ using gravitational lensing. These bound systems are candidate predecessors of today's globular clusters (GCs) \citep{vanzella_jwstnircam_2023, adamo_bound_2024, mowla_formation_2024, Fujimoto2024} and therefore continually shape our understanding of star formation in the early Universe.

Detailed numerical studies of star formation in massive ($\gtrsim 10^5~{\rm M_\odot}$) giant molecular clouds (GMCs) link high SFEs to regions of high surface density and pressure \citep[\eg][]{2018MNRAS.474.4232K, he_simulating_2019,  Fukushima2021, grudic_starforge_2021, lancaster_star_2021, Polak2024, 2023MNRAS.521.5160M, Menon2024}. These conditions are more prevalent at high redshifts, where massive star clusters that likely evolve into GCs today are formed \citep{shapiro_star-forming_2010, kruijssen_globular_2015, MBK2024-preprint}. Recent efforts focus on modelling their formation within a fully cosmological framework in the first galaxies \citep[\eg][]{garcia_star_2023, Rodriguez2023, Chen2023, Vanndonkelaar2023, Pfeffer2024, Calura2024arXiv}, with some focused on their roles as drivers of cosmic reionization \citep{ricotti_did_2002, katz_two_2013, Renzini2017MNRAS.469L..63R, Ma2021}.

Star cluster formation may also play a key role in the seeding and growth of supermassive black holes (SMBHs). JWST observations reveal the prevalence of compact ($\lesssim 200$ pc) high-redshift ($z \gtrsim 5$) objects with red continua, known as little red dots (LRDs) \citep{2023ApJ...956...61A, matthee_little_2024, 2024ApJ...964...39G}. Some exhibit line broadening (in some cases as high as FWHM $\sim 2000 ~{\rm km \: s^{-1}}$), suggesting masses of $10^{7-8} ~{\rm M_\odot}$ \citep{matthee_little_2024}. Whether these systems are powered by SMBHs or star formation remains uncertain \citep[\eg][]{Baggen2024}, highlighting the need to understand possible pathways for SMBH seeding.

Several mechanisms have been proposed to explain the rapid early growth of SMBHs, requiring either massive seeds, super-Eddington accretion, or both \citep[see review by][]{inayoshi_assembly_2020}. One commonly invoked scenario is the direct collapse black hole scenario where a metal-free, warm ($10^4$ K) primordial gas cloud avoids fragmentation and collapses into a $\sim 10^{5-6} ~{\rm M_\odot}$ seed \citep[\eg][]{Oh&Haiman2002, 2019Natur.566...85W}. However, these classes of models require rather specific conditions such as metal-free gas and strong irradiation by ultraviolet (UV) in the Lyman–Werner (LW) bands \citep[e.g.,][]{2019Natur.566...85W, James2025_preprint}.

Another established SMBH seeding mechanism involves runaway collisions in dense star clusters undergoing core collapse, particularly in the core of nuclear star clusters (NSCs), which are some of the densest known stellar systems with surface densities upwards of $10^6 ~ {\rm M_\odot \; pc^{-2}}$ \citep[see review by][]{neumayer_nuclear_2020}. NSCs can be triggered to collapse by sudden gas inflows from the surrounding galaxy \citep{Davies2011}, leading to the formation of an intermediate-mass ($\sim 10^{3-4}~{\rm M_\odot}$) SMBH seed via stellar collisions \citep{katz_seeding_2015} or black hole mergers \citep{Kritos2024_preprint}. \cite{Bellovary2025} recently suggested that LRDs may result from tidal disruption events in collapsing star clusters, connecting SMBH seed formation to star formation. Recent observations highlight the possibility of these scenarios: LRDs have distinct v-shaped spectral energy distributions (SEDs) that turn over at the Balmer limit, suggesting ongoing star formation \citep{2024ApJ...968...38K, Setton_vshaped}.

In \cite{garcia_star_2023} (hereafter \citetalias{garcia_star_2023}), we examined how different star formation efficiencies (SFEs) in star-forming clouds influence the formation, demographics, and survival of Pop II star clusters in a $z>6$ galaxy. Here, we build on that work by adopting an SFE prescription based on high-resolution simulations of gravoturbulent molecular clouds presented in \cite{he_simulating_2019}. Our goal is to track realistic star cluster formation in a typical high-$z$ dwarf galaxy -- of which are most responsible for reionizing the Universe \citep{Atek2024} -- and while doing so, outline a pathway for NSC formation. Future work will extend these simulations to explore SMBH seed formation in NSCs.

This paper is structured as follows: Section~\ref{sec:Methods} details our astrophysical models and numerical methods, Section~\ref{sec:Results} presents our findings, and Section~\ref{sec:Discussion} discusses their implications. We summarize our conclusions in Section~\ref{sec:Conclusions}.

\section{Methods and Simulations}\label{sec:Methods}

We use a branched version of the adaptive mesh refinement (AMR) code \textsc{ramses-rt} \citep[][]{Teyssier2002, rosdahl_ramses-rt_2013}. This radiative hydrodynamics (RHD) code employs a fast, moment-based approach to solve the radiative transfer equations on AMR grids using the M1 closure relation \citep{rosdahl_ramses-rt_2013}. Our version of the code contains astrophysical modules developed and implemented from past studies of reionization-era galaxies \citep{kimm_feedback-regulated_2017, katz_interpreting_2017}, star formation in isolated molecular clouds \citep{he_simulating_2019, he_simulating_2020}, and Pop III star formation \citep{ park_population_2021-1,park_population_2021, park_population_2023, park_origin_2024}. Most recently, we used this code to investigate the effects of UV feedback on the formation of Pop III stars in primordial halos and its transition to Pop II stars in the first galaxies \citep{sugimura_violent_2024}. Many of the numerical methods and physical models are identical to those introduced in \citetalias{garcia_star_2023} and \cite{sugimura_violent_2024}. Please refer to these papers for specific details of the initial conditions, gas chemistry, and cell refinement strategy. Here, we review key features of the simulations and highlight improvements made to our star-formation model. 

We run a cosmological ($h=0.70$, $\Omega_{\rm b} = 0.044$, $\Omega_{\rm m} = 0.27$, and $\Omega_\Lambda = 0.73$) zoom-in simulation on a dark matter (DM) halo that grows to become a dwarf galaxy with halo mass $M_{\rm h}\sim10^{10}~{\rm M_\odot}$ by $z = 0$. This halo is in a $(300 \: h^{-1} \: {\rm ckpc})^3$ refined region part of a larger $(35  \: h^{-1} \: {\rm cMpc} )^3$ box. We evolve this halo from initial conditions generated at $z = 127$ with \textsc{MUSIC} \citep{hahn_multi-scale_2011} to $z \sim 8.0$ (note, however, that this varies slightly depending on the star formation model used, see Section \ref{sec:Results}). Using AMR, we require all cells in our zoom-in region (with initial refinement level $l = 14$ and $800 ~ {\rm M_\odot}$ DM mass resolution) to be refined if they contain: (i) more than 8 DM or star particles or (ii) a gas mass exceeding $160~{\rm M_\odot}$, which is around eight times the initial mean gas mass per cell in the refined region. We also use a Jeans criterion; see \citetalias{garcia_star_2023} for additional details. Using this scheme, we achieve a maximum refinement level of $l =25$, corresponding to a minimum spatial resolution of $\Delta x_{\rm min} = 0.15[(1+z)/10]^{-1}$~pc. 

\subsection{Star formation and feedback}

We form stars in maximally refined ($l = 25$) cells when the gas density reaches the critical density threshold: 
\begin{equation}
    \begin{aligned}
    n_{{\rm H, crit}}=&\left(5.0\times 10^4 ~{\rm cm}^{-3}\right) 
    \left( \frac{T}{100~{\mrm{K}}} \right) \left(\frac{1+z}{10}\right)^2 \\
    &\times \left(\frac{N_{\rm cr}}{4}\right)^{-2},
    \label{eq:n_cr}
    \end{aligned}
\end{equation} 
where $T$ is the temperature and $N_{\rm cr}$ is a free parameter that ensures that the Jeans length $\lambda_{\rm J}$ is resolved with $N_{\rm cr}$ cells at the maximum refinement level: $\lambda_{\rm J}=N_{\rm cr} \Delta x_{\rm min}$. We adopt a fiducial value of $N_{\rm cr} = 4$ for our simulations. Once the density of any given cell $n_{\rm H} > n_{{\rm H, crit}}$, star formation is treated in a sub-grid manner depending on the metallicity $Z$ of the cell. 

\subsubsection{Star formation prescription} \label{sssec: star formation presc.}

We adopt a metallicity threshold $\rm{Z_{crit}}=10^{-5}$~Z$_\odot$ (with $\rm Z_\odot = 0.02$) that determines whether we form a Pop III system or Pop II star cluster. If $Z < Z_{\rm crit}$, we form a single particle with mass 120$~{\rm M_\odot}$, representing a Pop III binary system consisting of a 40 and 80 $~{\rm M_\odot}$ star. Otherwise ($Z \geq Z_{\rm crit}$), we form a Pop II star cluster centred on the cell, consisting of individual massive star particles with masses $m_* = 10 ~{\rm M_\odot}$. To do this, we identify all cells with $n_{\rm H} > n_{{\rm H, crit}}$ and then construct spherically-averaged one-dimensional (1D) gas density profiles on the fly, centred on the peak density $n_{\rm peak}$, which is analogous to a core density. The size of the cloud ($r_{\rm cloud}$) is defined as the radius at which the 1D profile reaches a cut-off density of $n_{\rm cut, cloud} \equiv N_{\rm cut}^{-1} n_{\rm peak}$, where we use a fiducial value of $N_{\rm cut} = 10$ in our simulation. Since we are monitoring all cells each time step, in most cases, a single cell (or at most a small number of clustered cells) meets the density threshold at a step. But if there are multiple spatially disconnected cells that reach $n_{\rm H} > n_{{\rm H, crit}}$, we randomly pick a single cell as the cloud center within a 100 pc region and start the profile from there, treating that as the cloud center.

From $r_{\rm cloud}$, we can determine the mass ($M_{\rm cloud}$), average density ($n_{\rm cloud}$), and average metallicity ($Z_{\rm cloud}$) of the cloud. Note that \cite{sugimura_violent_2024} adopts different thresholds due to lower resolution and convergence requirements. Within $r_{\rm cloud}$, we instantaneously form stars in regions with densities $> n_{\rm cloud}$ in a randomly distributed manner weighted by the gas density (i.e., a star is more likely to be formed in a denser part of the cloud), injecting them with a velocity equal to the gas velocity of the cell in which they form.

Since the star formation timescale is typically $\lesssim 1 ~{\rm Myr}$ in high-density molecular clouds \citep{he_simulating_2019}, characteristic of the high-$z$ Universe \citep{adamo_first_2024}. Moreover, thermal feedback from Type II supernovae (SNe) occurs on timescales of $\geq 4 ~{\rm Myr}$ and the gas densities in our clouds are high enough to be optically thick for radiation to penetrate and affect neighboring, unresolved stellar cores that form concurrently.

The star cluster formed has a mass of $m_{\rm star\: cluster} = f_* M_{\rm cloud}$, where we adopt a total star formation efficiency of the star-forming cloud:
\begin{equation}
    \begin{aligned}
     f_{*} =
     & \: 0.004
     \left(\frac{Z_{\rm cloud}}{10^{-3}~{\rm Z}_\odot}\right)^{0.25} 
     \left(\frac{M_{\rm cloud}}{10^{4}~{\rm M}_\odot}\right)^{0.4}\\  
     &\times \left(1+\frac{n_{\rm cloud}}{n_{0}}\right)^{0.91},
     \label{eq: SFE}
     \end{aligned}
\end{equation}
with a maximum value of $f_* = 0.90$ and $n_{0} = 100~{\rm cm^{-3}}$. This sub-grid model for SFE ($f_*$) is derived from a suite of high-resolution (AU-scale), radiative magneto-hydrodynamic (RMHD) simulations of star formation in isolated molecular clouds conducted by \cite{he_simulating_2019}. These resulting trends, while not substantially confirmed by galaxy surveys in the local universe \citep[e.g., PHANGS][]{2025ApJ...985...14L, Meidt25}, are in line with expectations from other simulations \citep[e.g., see][]{Grudic2022, Polak2024}.

Furthermore, we assume that all the stars formed in a given cloud have the same metallicity $Z_{\rm cloud}$ and therefore can be treated as a single stellar population (SSP). As we will discuss later in this paper (Section \ref{sssec: Multiple populations in star clusters}), this is a rather approximate treatment since molecular clouds are hierarchical structures that have inhomogeneous metallicities \citep{chevance_molecular_2020, mondal_dependence_2024}.

\subsubsection{Radiative and thermal feedback from stars}\label{sssec:Radiative and thermal feedback from stars}

Our simulations follow the evolution of radiation emitted from stars in four frequency bins: ${\rm H_2}$-dissociating far-UV (FUV) LW radiation ($11.2~\mathrm{eV}<h_\mathrm{p}\nu< 13.6~\mathrm{eV}$); \HI ionizing extreme-UV (EUV) ($13.6 ~{\rm eV}<h_\mathrm{p}\nu<24.6~\mathrm{eV}$); \HeI ionizing ($24.6\mathrm~{\rm eV}<h_\mathrm{p}\nu<54.4~\mathrm{eV}$); and \HeII ionizing ($54.4\,\mathrm{eV}<h_\mathrm{p}\nu<200\,\mathrm{eV}$) radiation. In the Pop III star systems, the 40 ${\rm M_\odot}$ star emits UV radiation for 4 Myr before undergoing a hypernova \citep{schaerer_properties_2002} which injects $E_{\rm SN, Pop III} = 3 \times 10^{52} ~{\rm erg}$ in thermal energy and releases ejecta with mass $M_{\rm ejecta, Pop III} = 20 ~{\rm M_\odot}$ (of which $M_{\rm metal, Pop III} = 9 ~{\rm M_\odot}$ are metals) into the surrounding gas. The 80 ${\rm M_\odot}$ star directly collapses, feedback-free, into a black hole \citep{wise_birth_2011}. Though the remnant of this Pop III binary can either form a BH binary or merge to form a single BH, both cases are numerically represented by a $100 ~{\rm M_\odot}$ BH particle. Although this particle produces no feedback in the form of X-ray radiation and jets, BH accretion -- which we model using the Bondi-Littleton-Hoyle formalism \citep{Bondi1952} -- can emit negligible UV radiation in our simulations. Furthermore, these BHs grow very little given that they are in low-density environments resulting from the hypernova explosion preceding their formation \citep{sugimura_violent_2024}. 

Although we do not directly sample an initial mass function (IMF) for the Pop II masses (recall that $m_* = 10 ~{\rm M_\odot}$), individual Pop II particles emit UV radiation based on their age and metallicity following the radiative yields for LW, H-ionizing, and He-ionizing radiation \citep{kimm_feedback-regulated_2017, katz_interpreting_2017} assuming a Salpeter IMF (1 - 100 ${\rm M_\odot}$) \citep{salpeter_luminosity_1955}. The Salpeter IMF is also used to calculate SNe yields from the star clusters, with each Pop II SNe (we expect around 1 per 100 ${\rm M_\odot}$ in stars) injecting thermal energy $E_{\rm SN, II} = 10^{51} ~{\rm erg}$ stochastically $4 - 40 ~{\rm Myr}$ since the formation of the star cluster \citep{leitherer_starburst99_1999}.   
Of course, there are caveats to the astrophysical models we have discussed thus far. For example, our current Pop III sub-grid model is rather simplistic: it does not capture the multiplicity, orbital architectures, and mass function that recent theoretical works have suggested for these first stars \citep[\eg][]{Sugimura2020, Costa2023, park_population_2023, Sugimura2023}. Furthermore, the Pop II star clusters consist of massive, $10 ~{\rm M_\odot}$ star particles which significantly impact their secular evolution. This choice, made due to computational cost, can affect the orbital relaxation of low-mass ($m_{\rm star-cluster} \lesssim 10^3~ {\rm M_\odot}$) star clusters and cause them to undergo core collapse and hence evaporate artificially early \citep[][]{Spitzer1987}. In addition, while we stochastically sample the thermal feedback for core-collapse SNe for the Pop II star clusters, recent studies suggest that different IMF sampling methods can affect the overall stellar mass yields as well as metal enrichment of the ISM in dwarf galaxies \citep{Applebaum2020, Jeon2024}. 

\section{Results}\label{sec:Results}

We present the results of a new simulation that builds on the high-SFE (HSFE, with constant $f_* = 0.70$) and low-SFE (LSFE, $f_* = 0.35$) runs first introduced in \citetalias{garcia_star_2023}. In our previous work, the two values for $f_*$ were chosen as lower and upper limit roughly bracketing values expected from \cite{he_simulating_2019}, based on the masses and densities of star-forming gas clouds in our simulations. This initial study was, in part, intended as a controlled numerical experiment to test the dependence of star cluster demographics on the choice of the sub-grid $f_*$. 

\begin{figure*}
    \includegraphics[width=\textwidth]{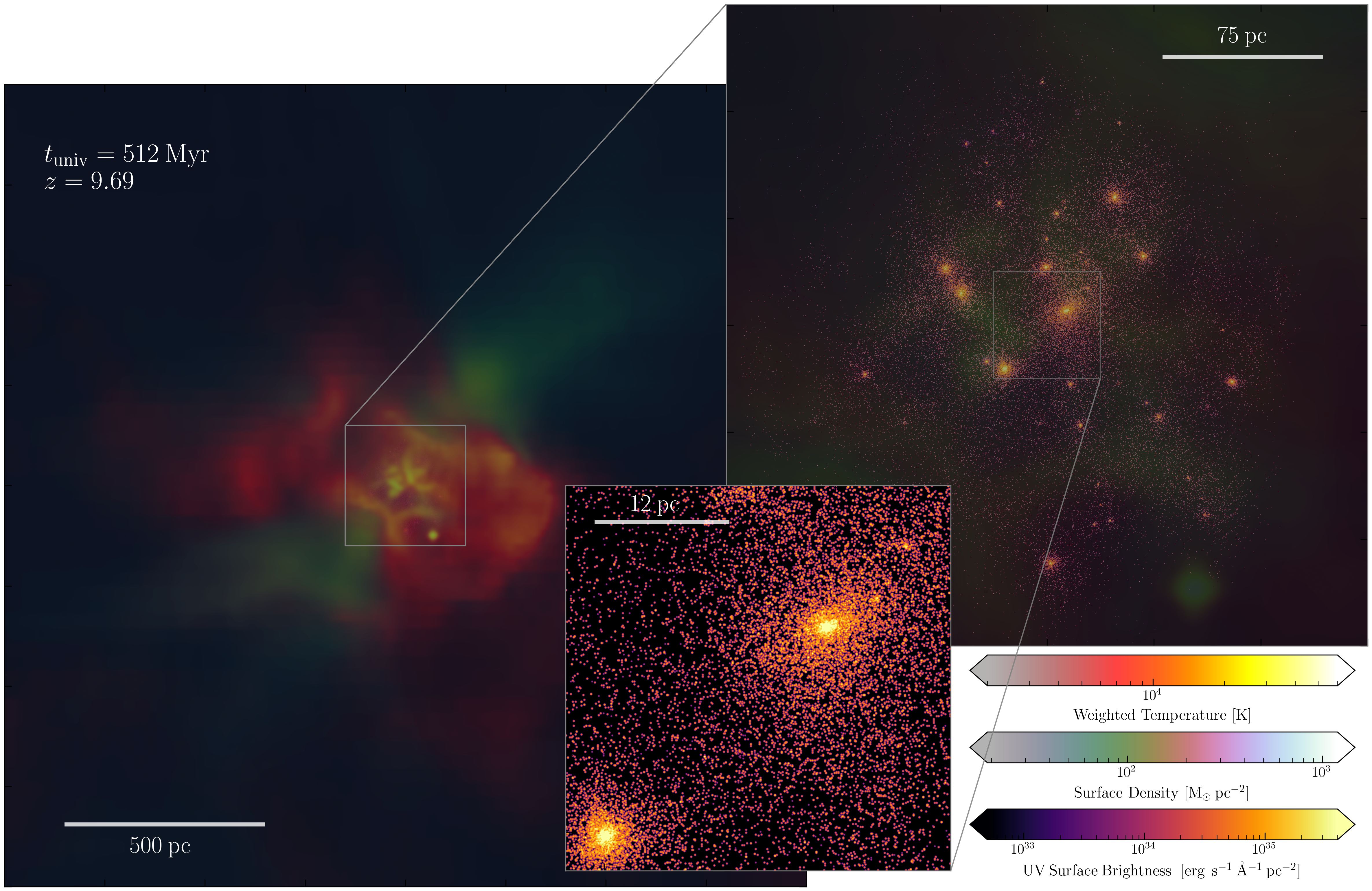}
     \caption{Snapshot of a dwarf galaxy simulation at high redshift with SFEs in gas clouds calibrated from cloud-scale simulations of star formation, shown just after starburst (a) in Figure \ref{fig: SF History}. The star clusters are shown according to their rest-frame UV surface brightness (at $\lambda = 1500 \text{\AA}$) assuming a Salpeter IMF and a metallicity of $10^{-3}~\rm{Z_\odot}$ using Starburst99. The green hues show the gas density while the redder hues show the density-weighted gas temperatures; the colour bars are at the bottom right of the figure. The panels progressively zoom into the star clusters (scale bars shown in physical units), with the bottom panel showing two of the most massive star clusters. The most massive star cluster, shown near the top right corner of the inset, comprises roughly 10\% of the galaxy's total stellar mass. }
     \label{fig: vsfe run render}
\end{figure*}

The new simulation presented here is not just a numerical experiment. Rather, it aims to reproduce, with the best fidelity, star formation in a typical dwarf galaxy at high-$z$ by adopting values for the $f_*$ derived from AU-scale RMHD simulations of star formation in molecular clouds \citep{he_simulating_2019}. In the new run (VSFE), the sub-grid SFE varies per cloud, depending on its metallicity, mass, and density (see Eq. \ref{eq: SFE} for the sub-grid $f_*$ model). Moreover, the HSFE and LSFE simulations analysed here incorporate updated data from continuing the simulations to lower redshift.

Figure \ref{fig: vsfe run render} shows a snapshot of the VSFE simulation at $z=9.69$, depicting the galaxy's star cluster population and gas properties about $20$~Myr after a starburst at $z \sim 10.2$\footnote{Animated figure available at \url{https://fred144.github.io/vids}}. The stars are shown by their post-processed rest-frame UV ($\lambda = 1500 ~\text{\AA}$) surface brightness using values from Starburst99 \citep{leitherer_starburst99_1999} tables assuming a Salpeter IMF between 1-100~M$_\odot$ and metallicity $Z=10^{-3}$~Z$_\odot$, similar to the procedure outlined in \citetalias{garcia_star_2023}. 

The first surprising result evident from the galaxy morphology in Figure \ref{fig: vsfe run render} is that the majority of the stars in the galaxy appear concentrated in a few compact, massive star clusters in the VSFE model -- even more massive and concentrated than those in the HSFE simulation (see Figure 12 in \citetalias{garcia_star_2023} for reference), which assumes that all stars form with a high SFE of 70\%. Later (Section~\ref{sssec: cmfs}), we quantify the star cluster mass function (CMF) slope and maximum mass. But qualitatively, these star clusters suggest that the VSFE run has a flatter power-law slope ($\Gamma \sim -0.5$) and higher truncation mass ($m_{\rm CMF,\: trunc} \sim 10^5$~M$_\odot$). Therefore when making a {\it physically motivated} assumption on the sub-grid SFEs in molecular clouds, we still find that galaxies at $z\gtrsim10$ {form \it most of their stars in bound star clusters}. Note that although not fully molecular, the star-forming clouds in this simulation serve as high-$z$ analogues to local molecular clouds and we use these terms interchangeably.

The analysis of simulation results is structured as follows. We begin by taking a look at the galaxy-wide star formation history in this VSFE run from $z \sim 12.8 - 8.0$, comparing it with the HSFE and LSFE in Section~\ref{ssec: SF history}. Then in Section~\ref{ssec: SFEs}, we analyse the metallicity and SFE evolution of the star-forming clouds throughout the simulations. Narrowing our focus to our VSFE model, we characterize the population of star clusters produced and their internal properties in Section~\ref{ssec: first clusters}. Finally, we present a possible formation scenario for an NSC at high-$z$ in Section~\ref{ssec: nsc seeding}.

\subsection{Bursty star formation histories} \label{ssec: SF history}

\begin{figure}
    \includegraphics[width=\columnwidth]{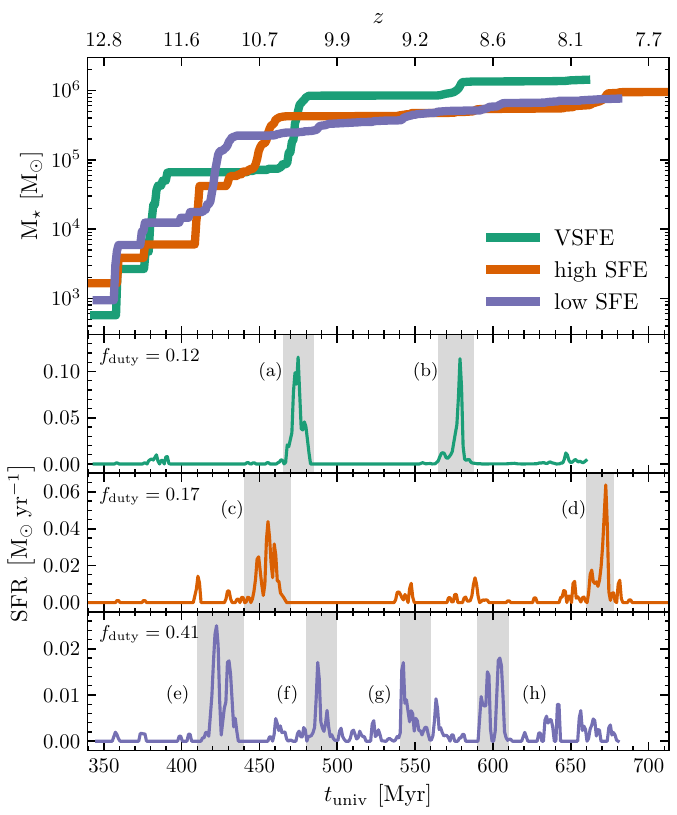}
    \caption{Pop II star formation histories of the simulations presented in this paper (see also Fig. 2 (left) in \citetalias{garcia_star_2023}). The top panel shows the mass of stars produced under the different assumed star formation efficiencies throughout each simulation (see legend). The following panels (top to bottom) show the corresponding star formation rates (same colours as the legend) in 1 Myr bins. In the top left of each panel, we show the duty cycle $f_{\rm duty}$ of star formation, defined as the ratio between times when the SFR $> 5$\% of the peak SFR, and the total time elapsed since the onset of Pop II star formation through the end of each simulation run. Note that we also label starbursts (a) - (h) across all simulations, shading regions where the SFRs roughly reach at least 5\% of the peak SFR during the starburst. We will use these in the text to refer to star-forming periods (\eg ~starburst (a) and (b) for the first and second starbursts in the VSFE model, respectively).}
    \label{fig: SF History} 
\end{figure}

Although the SFEs in the molecular clouds in the VSFE run are mostly $< 70$\% (see Section~\ref{ssec: SFEs}), the total mass in Pop II stars formed at $z < 10$ is nearly twice that of the HSFE run, as shown in the top panel of Figure~\ref{fig: SF History}. Star formation appears more stochastic, with longer quiescent periods followed by stronger bursts (Fig.~\ref{fig: SF History}, bottom three panels). We previously found in \citetalias{garcia_star_2023} that the total mass in stars at a given time -- and hence the mass of metals -- was nearly independent of the assumed sub-grid $f_*$ due to self-regulation \citep[\eg][]{ricotti_did_2002, Yajima2017}. However, by adopting the VSFE model, the total stellar mass formed in the galaxy varies. This suggests that the level of self-regulation depends non-trivially on the star formation and feedback recipe. 

Note, the DM halo mass of this galaxy is about $10^8$~M$_\odot$ at $z \sim 10$. This is a typical-mass galaxy at this redshift, having a stellar mass comparable to, or slightly lower than, the faintest galaxies observed by JWST at similar redshifts. A rarer, more massive halo would yield higher stellar masses and thus higher peak star formation rates during major bursts. However, here we focus on the physics of star formation by comparing the properties of the same galaxy when changing the sub-grid recipe for star formation.

The VSFE run produces the most mass in stars. Figure~\ref{fig: SF History} (top panel) shows that the VSFE, HSFE, and LSFE produced (Pop II) stellar masses $M_\star = 1.4 \times 10^6$, $ M_\star =9.5 \times 10^5$, and $M_\star = 7.6 \times 10^5~{\rm M_\odot}$, respectively. In the lower panels, we show the star formation rate (SFR) for each simulation sampled in 1 Myr bins and show that the VSFE simulation roughly doubles the peak SFR (${\rm SFR}\sim0.12 ~\rm{M_\odot~yr^{-1}}$ at 475 Myr) of the HSFE simulation (${\rm SFR}\sim0.06 ~\rm{M_\odot~yr^{-1}}$ at 672 Myr) and is four times higher than that of the LSFE simulation (${\rm SFR}\sim0.025 ~\rm{M_\odot~yr^{-1}}$ at 422 Myr). Note, the HSFE and LSFE star formation histories extend those initially depicted in \citetalias{garcia_star_2023}, Figure 2. For example, we show an additional starburst -- starburst (d) -- in the HSFE galaxy. 

We also calculate the duty cycle ($f_{\rm duty}$), defined as the ratio between the times when the galaxy SFR $> 5$\% of its peak SFR and the total time elapsed for Pop II star formation. The $f_{\rm duty}$ for the VSFE run is the lowest out of the three at $f_{\rm duty, VSFE} = 0.12$, as opposed to $f_{\rm duty, HSFE} = 0.17$ and $f_{\rm duty, LSFE} = 0.41$. Star formation is even more stochastic using the VSFE model than it is with the HSFE: it is more quiescent in between major starbursts, having fewer small bursts with ${\rm SFR} \gtrsim 0.01 ~\rm{M_\odot~yr^{-1}}$ (Figure \ref{fig: SF History}, grey regions) while also having the highest peak SFR. 

Following the arguments in \citetalias{garcia_star_2023} and \cite{sugimura_violent_2024} that suggest the burstiness of the star formation is feedback-mediated by FUV radiation, clouds that form after star forming episodes require higher masses for gravitational instability (\eg ~see \citetalias{garcia_star_2023} Eq. 4) and higher densities for cooling via self-shielding in a metal-poor environments. Adopting an SFE model that depends on the mass and compactness of the clouds has the effect of boosting starbursts resulting from a few massive clouds and suppressing smaller star-forming episodes arising from lower mass and density clouds. This allows the galaxy to accumulate a gas reservoir leading up to a SFB. This increase in mass and density leads to higher SFEs during starburst in our simulations (Eq. \ref{eq: SFE}). This non-linear effect -- which suppresses small bursts (in low mass or density clouds) and enhances larger bursts (high mass or density clouds) -- is reflected in both the star-formation histories and the total stellar masses. This is contrary to the assumption of a constant SFE that is independent of cloud properties.

The total mass in star forming clouds in the VSFE model is significantly larger than in the constant SFE models (in the LSFE model the cloud mass is roughly twice than in the HSFE model since the total stellar mass in the two models is nearly the same). Hence, even though the mean SFE in the VSFE model is less than the 70\% adopted in the HSFE model, the total stellar mass is still twice as large. The reason for a larger mass in star forming clouds is less certain, and may be due to feedback source clustering in higher density regions of the galaxy in the VSFE model.

On the other hand, a more trivial explanation is that in the VSFE model, most of the low-mass/low-density clouds form too few stars to contribute to self-regulation at the scales we currently resolve. That is, they contribute to the cloud and stellar mass budget but not to the feedback-regulating portion of it. In other words, when feedback is effective, it sets a maximum limit on the number of stars (more precisely, the number of massive stars producing UV radiation) in the galaxy. However, if the feedback is too weak in a subset of low-mass clouds due to currently unmodeled/unresolved pre-SNe feedback (e.g., stellar winds), these clouds can form stars without being self-limited, solely adding to the stellar mass budget. This effect is less important assuming constant SFE, because each cloud contributes to the mass in stars and UV feedback linearly with its gas mass. Nonetheless, the stochasticity of star formation observed here can have important implications on the observations of UV-bright objects at high-$z$ \citep{pallottini_stochastic_2023, kravtsov_stochastic_2024, Vikaeus2024MNRAS.529.1299V}.

\subsection{Star formation efficiency of clouds} \label{ssec: SFEs} 

\begin{figure}
    \includegraphics[width=\columnwidth]{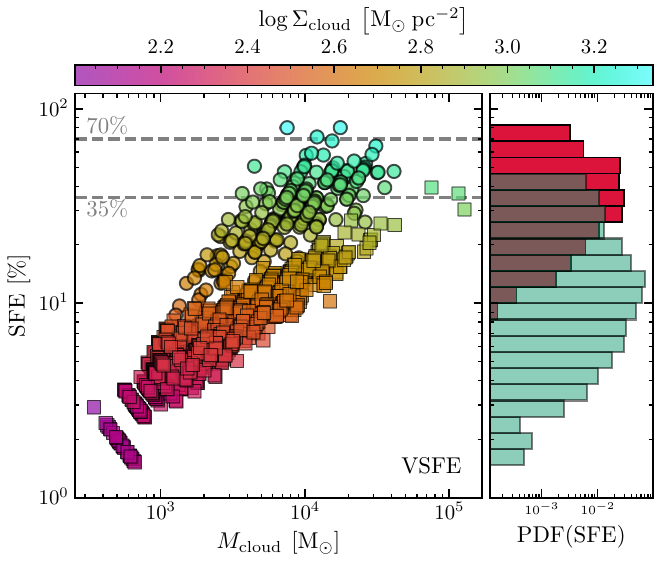}
    \caption{SFEs for the star-forming clouds in the VSFE simulation as a function of cloud mass. The \textit{circle} markers are clouds that produced star clusters before starburst (b) ($t \lesssim 500$ Myr) while the \textit{square} markers are for clouds that formed stars during and after starburst (b). The markers themselves are coloured by the mean cloud surface density $\Sigma_{\rm cloud}$ assuming a spherically symmetric, constant-density cloud. For comparison, we also show two dashed lines indicating the two constant efficiency runs: low SFE (35\%) and high SFE (70\%). The right panel shows the distribution of SFEs for these two time periods weighted by the stellar mass. The red histogram refers to the first burst (circle markers) and the green histogram refers to the second burst (square markers).} Note that the highest SFE achieved by the densest cloud is around 80\%, roughly 10\% higher than the high-efficiency fiducial run. 
    \label{fig: SFE}
\end{figure}

Figure~\ref{fig: SFE} shows the SFE of each cloud as a function of $M_{\rm cloud}$ in the VSFE simulation. The two simulations presented in \citetalias{garcia_star_2023}, which assumed cloud-scale star formation efficiencies to be 35\% and 70\%, are shown as dashed lines. Although the markers are coloured according to the cloud gas surface densities ($\Sigma_{\rm cloud} \approx M_{\rm cloud} / \pi r_{\rm cloud}^2$), recall that the cloud core densities can exceed $\Sigma_{\rm cloud}$. We also distinguish the points (and distributions) according to whether they are formed before (circles, red distribution) or during and after starburst (b) (squares, green). 

Looking at the stellar mass-weighted log probability distribution function (PDF) in the right panel of Figure~\ref{fig: SFE}, early star formation generally occurs at efficiencies between 10 to 80\% (with most of the stellar mass formed in clouds with SFE $\gtrsim$ 30\%) in clouds with surface densities reaching as high as $\Sigma_{\rm cloud} \sim 2.2 \times ~10^3 ~{\rm M_\odot \: pc^{-2}}$ ($n_{\rm cloud}\sim 4.8 \times 10^{4} ~{\rm cm^{-3}})$. We observe that the highest SFE cloud, although not the most massive throughout the simulation (at only $M_{\rm cloud} \sim 2 \times 10^{4}~{\rm M_\odot}$), is one of the densest and forms a $\sim 5 \times 10^3~{\rm M_\odot}$ star cluster at 80\% SFE during starburst (a). Similarly, we observe that three of the densest clouds form stars at efficiencies exceeding the 70\% efficiency assumed in the HSFE model. 

Cloud-scale SFEs before starburst (b) are reasonably bracketed by the LSFE and HSFE models, with roughly half falling in this range. However, the range of cloud SFEs during and after starburst (b) ranges between $\sim 2 - 40$\%, with the majority having efficiencies $< 35$\%, lower than what is assumed in the LSFE model. The SFEs are also closely tied to the SFRs observed in Figure \ref{fig: SF History}. In the HSFE model, the SFE is high even before reaching the peak of starburst (a), which has the consequence of suppressing star formation and reducing the peak SFR. Conversely in the LSFE model, the SFE is assumed to be low even at peak bursts reducing the peak SFR. 

A notable observation is that the most massive clouds ($M_{\rm cloud} \sim 10^5 ~{\rm M_\odot}$) form stars during starburst (b). This is consistent with the findings in \cite{sugimura_violent_2024} and the argument presented above in Section \ref{ssec: SF history}, where higher masses are needed for clouds to become gravitationally unstable and form stars at low metallicities and high temperatures from preceding star formation, which provide thermal support.

\subsubsection{Metallicity evolution of star-forming clouds} \label{sssec: Metallicity evolution of star-forming clouds}

\begin{figure}
    \includegraphics[width=\columnwidth]{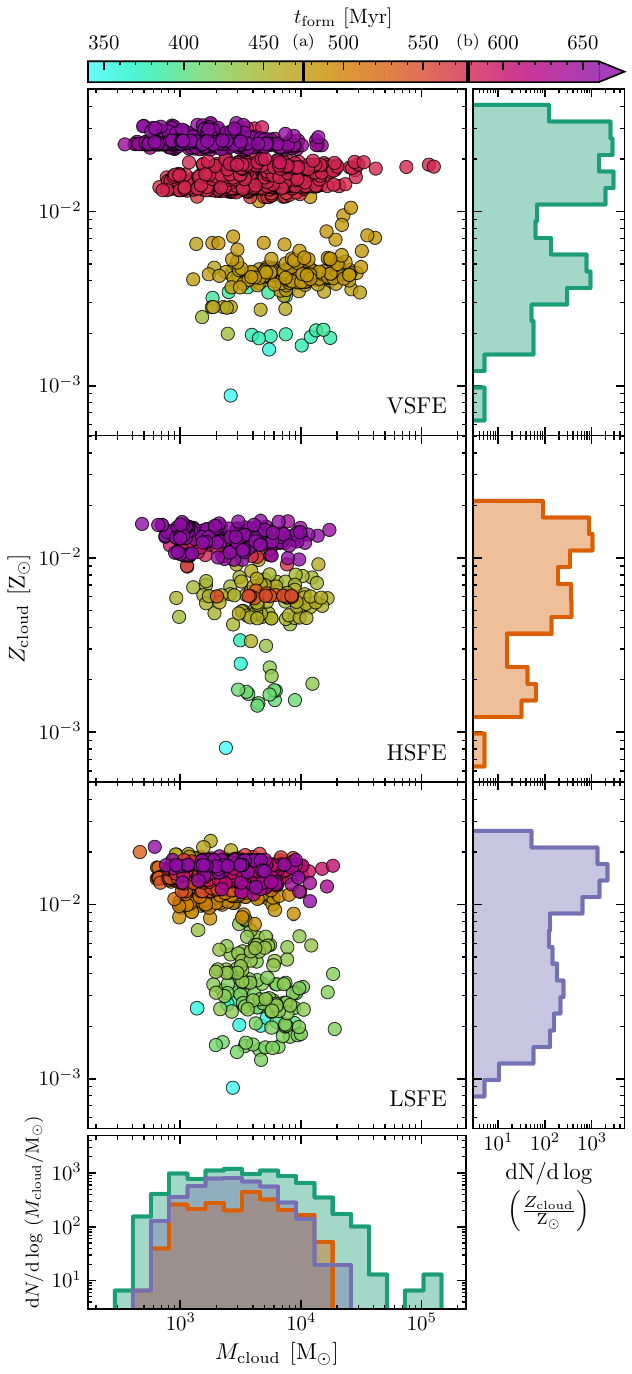}
    \caption{We show the Pop II star-forming cloud metallicity as a function of mass for the different star-forming periods: see top colour bar where we also indicate (with two thicker lines) the times when the SFRs peak for staburst (a) and (b). For each row, the right panel shows the metallicity function for the clouds for a given model (see lower right of each left panel). The very bottom panel shows the mass function of the clouds, coloured according to the sub-grid SFE for each run. Note, that the metallicity of the star clusters is identical to their natal clouds'.}
    \label{fig: cloud metals}
\end{figure}

The trends observed in Figure~\ref{fig: SFE} are partly explained by the metals produced by SNe in starburst (a), which enrich the ISM and lead to higher metallicity clouds during starburst (b). This becomes clearer when we turn our attention to Figure~\ref{fig: cloud metals}, which shows the relationship between $Z_{\rm cloud}$ and $M_{\rm cloud}$ for each simulation (see the bottom right of the left panels for each corresponding run). The right panels show the shape of the metallicity function: $\mathrm{d}N / \mathrm{d}\log (Z_{\rm cloud} / {\rm Z_\odot}$) at the end of each simulation, while the very bottom row shows the mass function of the clouds in all three simulations overlaid on top of another. We also colour each marker by the star cluster formation time in the simulation. Recall from Section \ref{sssec: star formation presc.} that the metallicities of individual Pop II stars formed are equal to $Z_{\rm cloud}$. Hence, $Z_{\rm cluster} = Z_{\rm cloud}$ in many cases -- especially at the formation of a Pop II star cluster. However, a cluster may contain multiple populations due to mergers. Also, this is not to be confused with $Z_{\rm Pop II}$ which is the metallicity of \textit{individual stars} and may not necessarily belong to the same (or any) star cluster throughout the simulation.

For all simulations, the lowest metallicity cloud ($Z_{\rm cloud} \break \sim8\times10^{-4}~{\rm Z_\odot}$) forms at $z = 12.95$ ($t_{\rm univ} \sim 350 ~{\rm Myr}$). This cloud forms a star cluster at $f_* = 0.22$ in the VSFE simulation, indicating that $f_* = 0.35$ and $f_* = 0.70$ both overestimate the efficiency of this cloud. Feedback from this first star cluster heats the surrounding gas, delaying subsequent star-forming periods. And since hotter clouds with negligible metal cooling require larger masses to become gravitationally unstable (Section \ref{ssec: SF history}), we get a succeeding population of more massive clouds that garner higher SFEs, reaching as high as 80\% in our VSFE model. We observe this in all simulations with cloud masses typically reaching around $2 \times 10^4 ~ {\rm M_\odot}$ after ($\lesssim 20$ Myr) the first star cluster forms, see Figure \ref{fig: cloud metals}. The initial star formation and metal enrichment histories are fairly similar across simulations; adopting the VSFE model has trivial consequences for the star formation immediately following the first star cluster. 

However, these physically motivated SFEs do play an important role during and after starbursts. Namely, they can significantly alter the total stellar masses, which are directly related to a galaxy's metal production and stellar feedback budgets. These feedback processes impact the next generation of star formation. Recall from Figure \ref{fig: SFE} that a significant portion of star formation in the VSFE model during starburst (a) occurs in higher SFE ($f_* \gtrsim 0.35$) clouds due to the compounding feedback from earlier star formation, progressively increasing the SFEs leading up to the burst. Because of this, clouds can reach SFEs as high as 80\% during starburst (a). The intense feedback from this high SFE burst leads to complete quiescence which we can see by the large gap ($\gtrsim 50$ Myr) in $t_{\rm form}$ for the top panel in Figure \ref{fig: cloud metals}. As we saw in \citetalias{garcia_star_2023}, more bursty star formation histories exhibit an evident multi-modal metallicity distribution leading to the LSFE model having a comparatively more gradual metal enrichment than the HSFE. Our results here are consistent with this, with the most stochastic model of the three ($f_{\rm duty, VSFE} = 0.12$) also having the clearest multi-modality (Figure \ref{fig: cloud metals}, right histograms). In addition to this multi-modality, the VSFE model reaches the highest metallicity among the three runs, despite not evolving this model down to lower redshifts (Figure \ref{fig: SF History}). 

We also find that the metallicity evolution of the star clusters is non-monotonic; \ie lower metallicity clouds may condense in the ISM even after higher metallicity star clusters have already formed. For example, in the VSFE model (Figure \ref{fig: cloud metals}, top row), there are lower metallicity clouds (orange points, $\sim 3 \times 10^{-3}~{\rm Z_\odot}$) that form \textit{after} clouds with higher metallicities (light green) had already formed stars. This also occurs later during and after the second starburst and across the other two models, possibly hinting at metallicity dilution via inflows of more pristine gas \citep[see,][]{sugimura_violent_2024, Stiavelli2024_preprint}, metal ejection via feedback, or metallicity inhomogeneities in the ISM before the metals are thoroughly mixed. In the LSFE simulation's case, the larger spread in metallicities around generations of star clusters (e.g., green points) is likely related to the less bursty -- closer to the continuum mode -- nature of star formation. Less intense star formation leads to a more gradual build-up of metals that get reincorporated and recycled almost immediately for the next generation of stars, as opposed to a more prolonged period of quiescence for the comparatively more bursty HSFE and VSFE models. 

The comparatively more intense heating from starburst (a) in the VSFE run leads to the formation of the most massive clouds (bottom row of Figure \ref{fig: cloud metals}, green) across all models during the following starburst (b), reaching masses of $M_{\rm cloud} \gtrsim 10^5 ~{\rm M_\odot}$. In comparison, the HSFE and LSFE reach cloud masses of only $M_{\rm cloud} \sim 2 \times 10^4 ~{\rm M_\odot}$. The general form of the cloud mass function is roughly consistent with a piecewise lognormal and power law, characteristic of a gravoturbulent star-forming environment \citep{burkhart_star_2018}. The dynamical evolution of the stars produced by these high mass, metal-enriched, lower-density (see Figure \ref{fig: SFE}, colour bar) clouds is discussed in further detail in Section \ref{ssec: nsc seeding}.

In summary, adopting the VSFE model alters the chemical enrichment history of star-forming clouds by enabling comparatively higher metal yields and thermal feedback budgets than the HSFE and LSFE runs. This allows the formation of higher-temperature, more massive clouds. However, these massive clouds harbour higher metal enrichment which can lead to lower SFEs ($f_* \lesssim 0.35$). Higher metal enrichment can also cause clouds to fragment to lower masses, leading to a transition to low SFEs we see in Figure \ref{fig: SFE} post starburst (b).

\begin{figure*}
    \includegraphics[width=\textwidth]{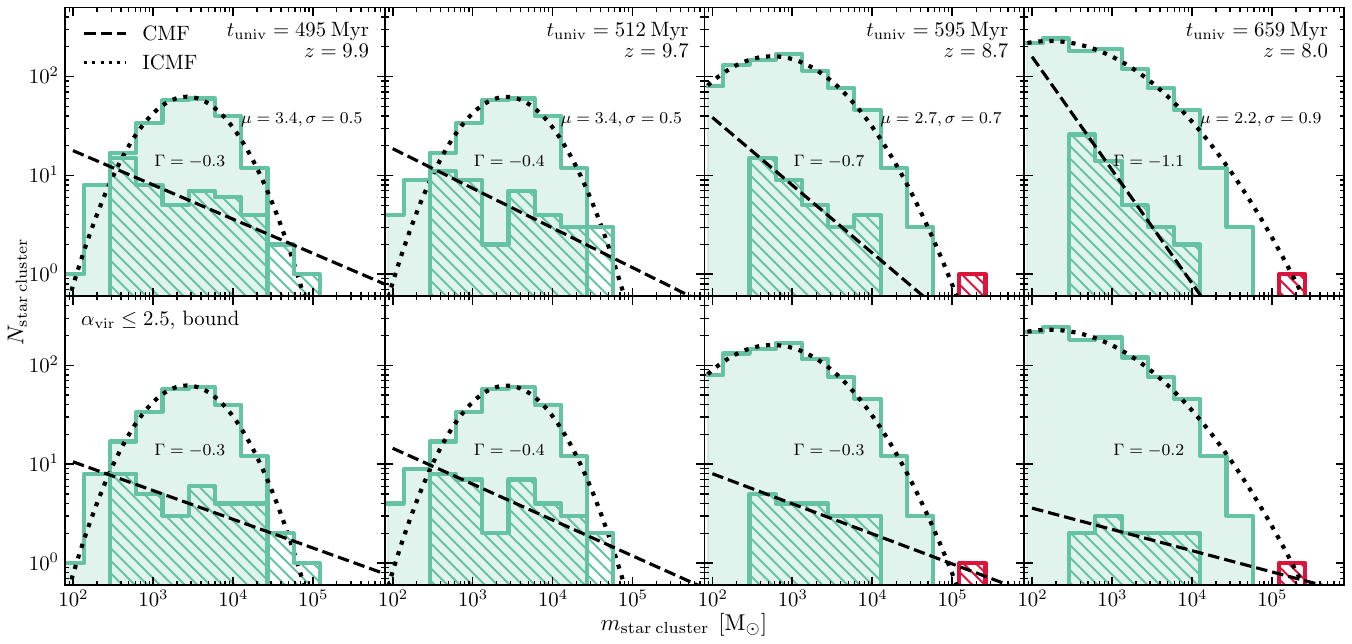}
    \caption{At-birth (solid) and current (hatched) star cluster mass function for the VSFE run. The simulation progresses from left to right and we see how the star cluster population evolves after starburst (a) from the first column and second column ($t_{\rm univ} = 512$ Myr corresponds to the snapshot shown in Figure \ref{fig: vsfe run render}),  right after the second starburst (the third column), and the latest snapshot (fourth column). We also show reference lines for power laws used to fit the CMF with power law index $\Gamma$ which also evolves. Note that the NSC seeded by the starburst (b) is indicated in red. Furthermore, we show the cumulative ICMF fitted with a Gaussian in log-log space to find the mean ($\mu$) and standard deviation ($\sigma$). The top row corresponds to all the star clusters, both bound and unbound (see text for definition), while the second row only shows those we consider bound.}   
    \label{fig: cmf} 
\end{figure*}

\subsection{The first star clusters} \label{ssec: first clusters}

So far, we have focused on the overall star formation histories across all simulations. In this section, we will shift our attention to the VSFE model, characterizing the properties of star clusters in the first 700 Myr.
\newpage
To identify star clusters, we post-process our snapshots with a friends-of-friends (FoF) structure finder \citep{efstathiou_numerical_1985}, grouping stars that are closer than $l_{\rm link} = 10^{-4}$ times the mean separation of all the star particles. For a given star cluster we calculate the total kinetic energy $E_{\rm kin}= \frac{1}{2} \sigma^2_{\rm 3D} m_{\rm star\: cluster}$, where $\sigma^2_{\rm 3D}$ is the 3D velocity dispersion of the stars, $m_{\rm star\: cluster}$ is the mass of the Pop II star cluster. We then approximate each clump as a uniform density sphere having total potential energy $E_{\rm pot} = - \frac{3}{5} G m^2_{\rm star\: cluster} r^{-1}_{\rm half}$, where $r_{\rm half\: mass}$ is the half-mass radius and $G$ is the gravitational constant. Knowing that for an object in virial equilibrium: $2 E_{\rm kin} = - E_{\rm pot}$, we can estimate the virial parameter $\alpha_{\rm vir} \sim - 2 E_{\rm kin} / E_{\rm pot}$. Similar to the commonly used virial parameter for molecular clouds, a cluster found by the FoF is considered to be virialized if $\alpha_{\rm vir} \leq 1$ and bound if $\alpha_{\rm vir} \leq 2$ \citep{kauffmann_low_2013}.

However, given that many of our star clusters (especially low mass ones) are not very well resolved due to their sizes being 1-2~pc and the gravitational softening $\Delta x_{\rm min} = 0.15$ pc, clumps may be considered as bound star clusters even with $\alpha_{\rm vir}$ slightly larger than 2. Furthermore, we require a clump to have a mass $\geq 300 ~{\rm M_\odot}$ (corresponding to at least 30 star particles) to be considered as a star cluster.

\subsubsection{Star cluster mass functions} \label{sssec: cmfs}

We find that the Pop II star CMFs in the VSFE model are well represented by a power law ${\rm d}N/{\rm d}\log M \propto M^\Gamma$ with slopes varying between $ \Gamma \sim -0.3$ to $-1.1$ depending on the time since a major starburst and whether or not we include unbound star clusters. Note that our definition of the power-law slope differs from the other often used in the literature ${\rm d}N/{\rm d} M \propto M^\beta$. Hence, $\beta\equiv\Gamma-1$ and after a major starbust we observe $\beta \sim -1.3$, that is flatter than the typical values $\beta\sim -2$ observed in galaxies at redshifts $z<6$. This result is in good agreement with a detailed analysis of the star cluster mass function in the strongly lensed galaxy at $z \sim 10$ dubbed "Cosmic Gems" \citep{Vanzella2025}. Figure~\ref{fig: cmf} (top row) shows the CMFs for all star clusters identified on-the-fly (bound and unbound) as hatched histograms and the initial cluster mass functions (ICMFs) as solid histograms. The ICMF is simply the at-birth CMF: $f_* M_{\rm cloud}$; \ie~ it is the distribution of star cluster masses (prior to and during $t_{\rm univ}$ in each column) produced from a given star-forming cloud, before any dynamical evolution.

The ICMF of the VSFE run (Figure ~\ref{fig: cmf}, solid distributions) is well described by a log-normal distribution: $f(M_{\rm cloud}) = a \exp \left[ (M_{\rm cloud} - \mu)^2 / (2 \sigma^2) \right]$, with fitted mean ($\mu$) ranging between 2.2 and 3.4, standard deviation $\sigma \sim 0.50 - 0.90$, and some normalization $a$. Given the definition of the ICMF, we expect its shape and extent to be similar to the distribution of $M_{\rm cloud}$ depicted in the bottom panel of Figure \ref{fig: cloud metals}. However, recall that the few massive ($\sim 10^5 ~{\rm M_\odot}$) star-forming clouds shown in \ref{fig: cloud metals} have relatively low SFEs ($\lesssim$ 35\%), hence the star clusters produced have maximum at-formation masses of a few $10^4~{\rm M_\odot}$. 

The general shape of the ICMF remains lognormal between the VSFE, LSFE, and HSFE simulation (note, we show a similar plot to Figure \ref{fig: cmf} in \citetalias{garcia_star_2023} Figure 8). However, the value of $\sigma$ tends to be larger for the VSFE as opposed to the HSFE and LSFE simulations, which range in $\sigma \sim 0.2 - 0.3$. Furthermore, the mean values for the ICMF are slightly lower at $\mu \sim 2$ compared to $\mu \sim 3$ by the end of the simulation. We attribute this to the fact the since LSFE and HSFE have constant SFEs, they can overestimate the low mass end of the ICMF and underestimate the high mass end -- whereas the VSFE model is able to capture the full range of SFEs and thus reflecting this in the ICMF.

The CMF is time-dependent due to the dynamical evolution of the star clusters (Figure \ref{fig: cmf}, left to right). For example, star clusters as massive as $10^5 ~{\rm M_\odot}$ are formed via mergers since the ICMF shows a maximum at-birth star cluster mass of $\sim 2 \times 10^4 ~{\rm M_\odot}$ after the high-efficiency starburst (a) at $z=9.9$. For reference, this star cluster is the main star cluster we zoom into in Figure \ref{fig: vsfe run render}. These massive clusters then lose mass as shown in the second column ($z=9.7$). In the third and fourth columns ($z = 8.7 - 8.0$), the CMF steepens indicating the formation of predominantly low-mass, unbound star clusters after the low-SFE starburst (b). This is confirmed when filtering for star clusters with $\alpha_{\rm vir} \leq 2.5$ (which includes those marginally bound) in the bottom row of Figure \ref{fig: cmf}. This filtering only slightly changes the slope of the CMF after starburst (a) (first two columns); however, it significantly alters them post-starburst (b), changing the slope from $\Gamma = - 0.7$ and $- 1.1$ to much flatter values of around $\Gamma = -0.2$ (bottom right of Figure \ref{fig: cmf}). We can also indirectly see this effect by looking at the mean mass of the ICMF, which decreases between the middle left ($\mu = 3.4$, $t_{\rm univ} = 512$ Myr) and middle right ($\mu = 2.7$, $t_{\rm univ} = 595$ Myr) panels. This is caused by low-efficiency star formation in highly-fragmented, metal-enriched, and low-mass clouds (Figure \ref{fig: cloud metals}). These star clusters have at-birth masses of a few $100 ~{\rm M_\odot}$ and quickly get disrupted.

\subsubsection{Seeding of a nuclear star cluster} \label{sssec: Seeding of a nuclear star cluster}

\begin{figure}
    \includegraphics[width=\columnwidth]{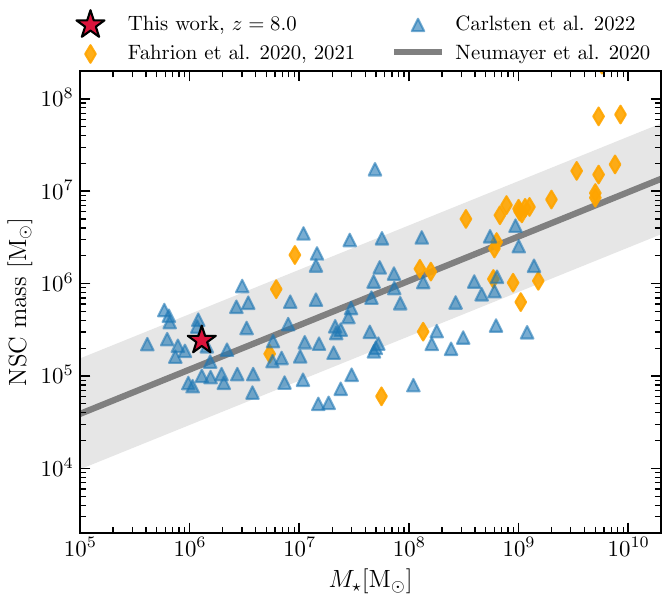}
    \caption{Nuclear star cluster mass as a function of host galaxy stellar mass. We compare the properties of the VSFE galaxy at $z = 8$ (red star) with those compiled for nucleated dwarf galaxies (at $z \sim 0$) from the Multi Unit Spectroscopic Explorer (MUSE) Fornax3D survey \citep[][]{Fahrion2021, Fahrion2022} as well as the Exploration of Local VolumE Satellites (ELVES) survey for low-mass early-type galaxies in the Virgo cluster \citep{carlsten_elves_2022}. For comparison, we show the scaling relation presented in a review by \protect\cite{neumayer_nuclear_2020} (Eq. 1), derived for a sample of 407 NSC and host galaxy masses $M_\star = 10^{6-11}~{\rm M_\odot}$ along with the scatter around the fit. }   
    \label{fig: nsc mass gal} 
\end{figure}

\begin{figure*}
    \includegraphics[width=\textwidth]{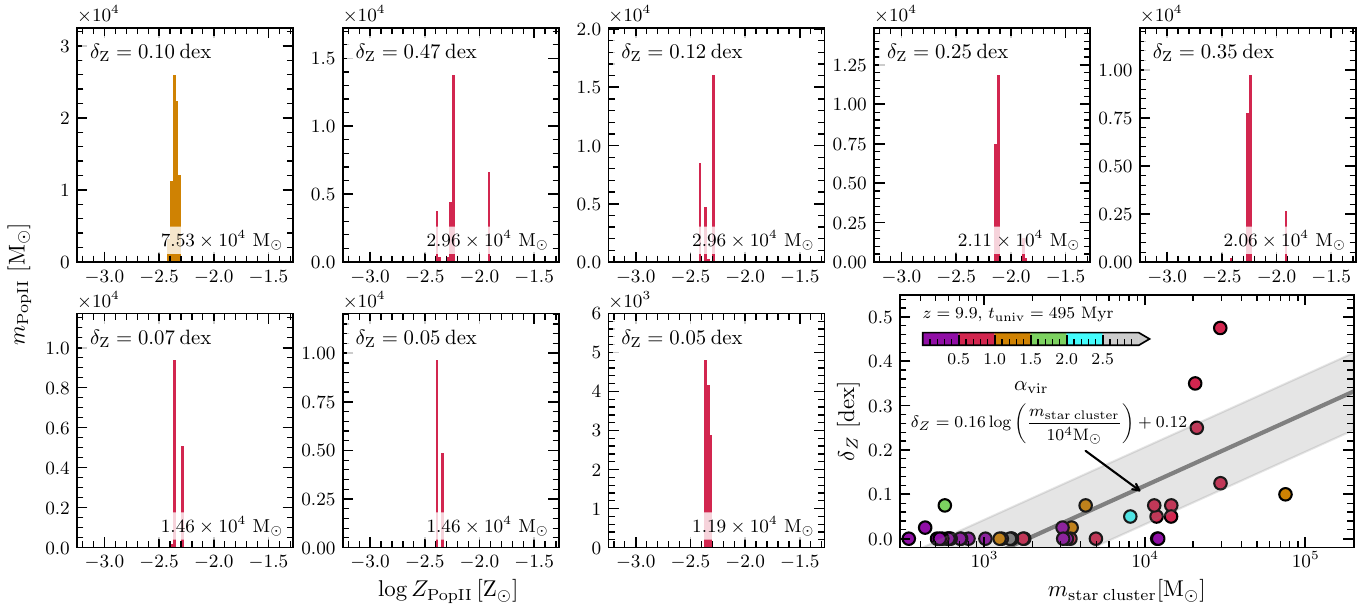} 
    \caption{Metallicity distribution of Pop II stars in the 8 most massive star clusters in our physically motivated (VSFE) simulation at $z = 10$, immediately after starburst (a). Each panel shows the distribution of Pop II star particle metallicities in a given star cluster (the total cluster mass is shown in each panel), starting with the two most massive in the top left (then across). The distributions are coloured according to $\alpha_{\rm vir}$ of the star cluster (see colour bar at the bottom right) and are displayed in log-linear space with 0.025 dex bins. We quantify the metallicity spread in each star cluster ($\delta_Z$) by calculating the range of metallicities from bins with appreciable mass contributions ($> 10$\% of the maximum mass in any given bin). In the bottom right panel, we show $\delta_Z$ as a function of star cluster mass for all identified star clusters, with points coloured by $\alpha_{\rm vir}$. We also show a rough log-linear fit for clusters with $\delta_Z > 0$ and $m_{\rm star \:cluster} \geq 10^4 ~{\rm M_\odot}$ 
    }   
    \label{fig: cluster zdist} 
\end{figure*}

Coincident with the second burst of star formation is the formation of a $\sim 10^5 ~{\rm M_\odot}$ object we highlight in red in the CMFs of Figure \ref{fig: cmf}. This object -- which we characterize as an NSC  -- is grown predominantly through the merger of star clusters formed in lower SFE clouds (see Section \ref{ssec: nsc seeding} for a more thorough examination). The initial growth of this object was first facilitated by a $5 \times 10^4 ~{\rm M_\odot}$ star-forming cloud during starburst (b). 

We further verify the existence of an NSC in Figure~\ref{fig: nsc mass gal}. The NSC has a mass of $m_{\rm NSC} = 2 \times 10^5 ~{\rm M_\odot}$ in a $1.4 \times 10^6 ~{\rm M_\odot}$ galaxy by $z=8.0$. For comparison, we depict nearby nucleated dwarf galaxies in the Fornax and Virgo clusters in the nearby ($z\sim 0 $) Universe from the MUSE Fornax3D \citep[][]{Fahrion2021, Fahrion2022} and the ELVES survey \citep{carlsten_elves_2022}. We also depict the NSC-to-galaxy stellar mass scaling relation presented in a review by \cite{neumayer_nuclear_2020}, where the NSC masses were found to scale as $m_{\rm NSC} \propto M_\star^{1/2}$. This indicates that NSCs contain a higher fraction of their host galaxy's mass in lower-mass galaxies. Indeed, the NSC formed in this work hosts roughly 20\% of the host galaxy's stellar mass by $z = 8.0$, falling within the 0.6 dex scatter around the expected scaling relation in Figure \ref{fig: nsc mass gal}.

\subsubsection{Multiple populations in star clusters} \label{sssec: Multiple populations in star clusters}

Due to the $m_* = 10 ~{\rm M_\odot}$ stellar mass resolution in our simulation, we can roughly trace how the dynamics of star clusters determine their internal stellar populations. Figure \ref{fig: cluster zdist} looks at this in closer detail by showing the metallicity distributions of Pop II star particles in the most massive star clusters at $z=10$, right after starburst (a). Note that each panel shows the metallicity distribution within a given cluster (with total mass shown in the bottom right).

The most massive star cluster ($m_{\rm star \: cluster} = 6.7 \times 10^4 ~{\rm M_\odot}$) has a noticeable 0.1 dex scatter in its metallicity distribution, with stars predominantly having metallicities between $ -2.4 < \log (Z_{\rm Pop II} / {\rm Z_\odot }) < -2.3$  and a negligible population (less than 1\% of the total mass) of lower-metallicity stars. We quantify the spread in a star cluster's metallicities ($\delta_Z$) by binning the log metallicities (0.025 bin sizes), masking out bins with counts $< 10$\% of the peak number of star particles in any given metallicity bin (this prevents a relatively small population from artificially increasing the calculated spread), and then taking the range of values. We observe that the most massive star clusters in Figure \ref{fig: cluster zdist} have $\delta_Z \sim 0.10 - 0.47$ dex, while clusters with $m_{\rm star \: cluster} \lesssim 10^4$~M$_\odot$ have less noticeable metallicity spreads $\delta_Z \lesssim 0.1$. Figure \ref{fig: cluster zdist} bottom right shows $\delta_Z $ for all identified star clusters. There is a weak trend between $\delta_Z$ and $m_{\rm star\: cluster}$, best described by the fitted log-linear relationship $\delta_Z = 0.16 (m_{\rm star\:cluster} /10^4 ~{\rm M_\odot}) + 0.12$ we show in the panel, along with a $\pm 0.08$ dex uncertainty derived from fitting. 

However, we caution the reader that the scatter we present here is spread across chemical species since we currently do not track individual elemental abundances and enrichment pathways (\eg~ AGB stars and Type Ia SNe); much higher model fidelity is needed to address questions about the origins of multiple stellar populations in GCs \citep{charlie_formation_2010, bastian_multiple_2018, bekki_formation_2019, el-badry_formation_2019}. However, this spread suggests that the metallicities of star-forming cloud complexes are non-homogenous and are already patchy even at sub-parsec scales and within the first Gyr of metallicity evolution within galaxies. While these star clusters are formed as an SSP, star formation within a cloud is hierarchical and sub-clumps of slightly different metallicities merge early on to form bound star clusters (the distributions are coloured according to the $\alpha_{\rm vir}$). While the dependence of metallicity spreads on the star cluster mass has been observed in local GCs with masses $\sim 10^{5-6} ~{\rm M_\odot}$ by \cite{Latour2025arXiv250109558L}, they can be interpreted in models that include self-enrichment \citep[\eg see][]{Bailin2018ApJ...863...99B, Madeleine}. However, if the star cluster is very compact and not too massive (as in our simulation), we expect that the timescale for self-enrichment is longer than the star formation quenching timescale \citep[$<2$~Myr,][]{he_simulating_2019}. The current metal enrichment scheme we use in our simulation forms stars in each gas clump instantaneously, and only tracks the enrichment of Type II SNe species, dominated by $\alpha$ elements and some Fe production \citep{Woosley1995, roberti_-process_2024}. Therefore, we interpret the spread seen here as due to inhomogeneities in the gas pre-enrichment rather than self-enrichment. This non-monolithic, hierarchical view of star cluster formation and assembly has been studied at the molecular cloud scale \citep[\eg][]{Vazquez2017, Grudic2022} and here we see a confirmation of this process at galactic scales in the ISM of high-$z$ galaxies. 

\begin{figure*}
     \includegraphics[width=\textwidth]{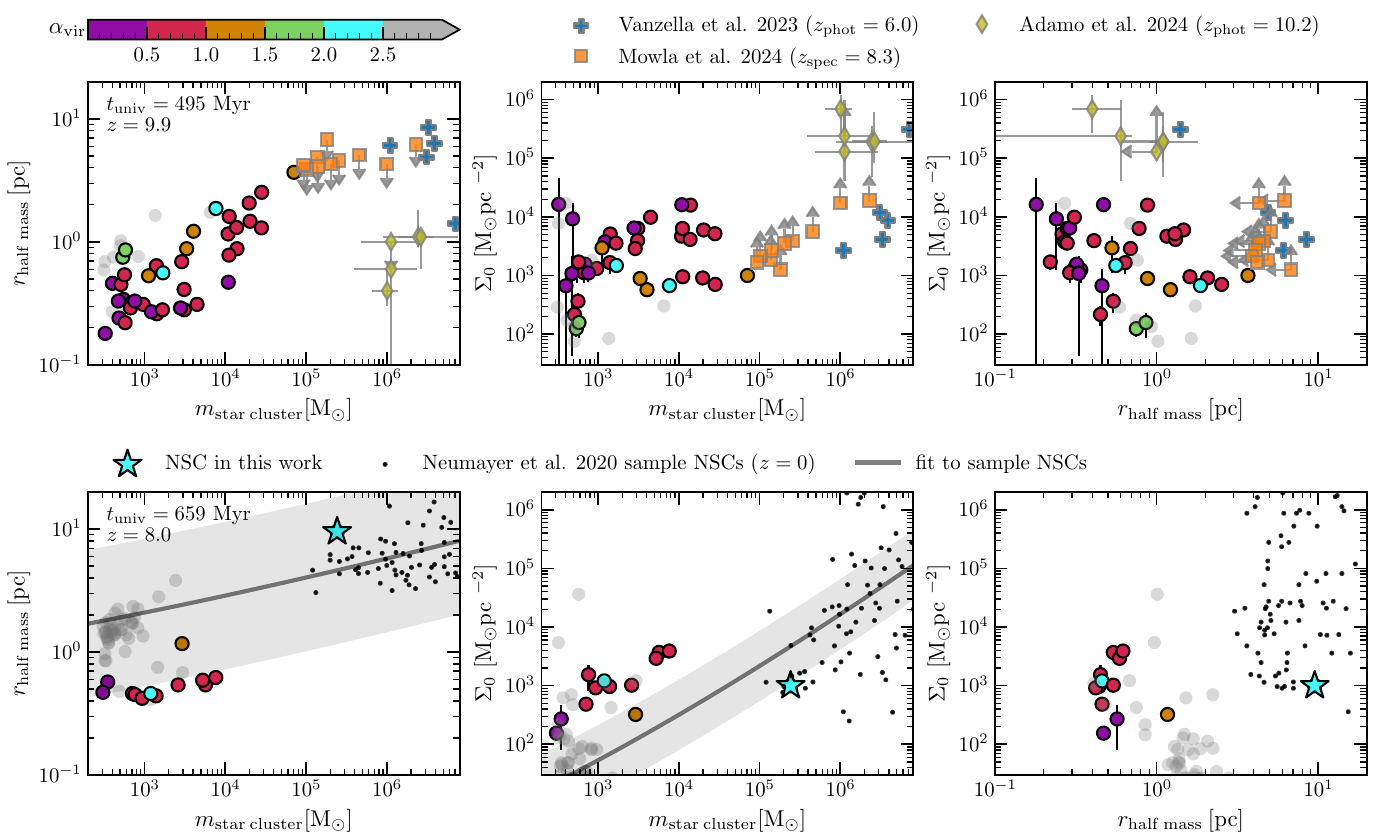}
    \caption{The evolution of the mass-radius relation (left column), mass-surface density relation (centre), and radius-surface density relation (right) of star clusters in the VSFE galaxy after starburst (a) ($t_{\rm univ}$ = 495 Myr, $\lesssim 20 ~{\rm Myr}$ since the peak of the burst) and at the end of the simulation ($t_{\rm univ}$ = 659 Myr, bottom row). The circles are coloured by the virial parameter if marginally bound (has a value of $\alpha_{\rm vir} \leq 2.5$) and grey otherwise. We also include properties of star clusters from the Sunrise Arc \citep{vanzella_jwstnircam_2023}, Cosmic Gems Arc \citep{adamo_bound_2024}, and Firefly Sparkle \citep[][]{mowla_sparkler_2022, mowla_formation_2024} observed through strong lensing at high-$z$ as points of comparison. In the bottom rows, we indicate the NSC formed and the surviving star cluster population at the end of the simulation along with a select sample of NSCs in early and late-type galaxies compiled in a review by \protect\cite{neumayer_nuclear_2020}. We fit the mass-radius and mass-density relation and show the results along with a 0.6 dex scatter (grey band) for reference.}  
    \label{fig: size mass relation} 
\end{figure*}

\subsubsection{Star cluster masses and sizes} \label{sssec: star cluster pop}

Figure \ref{fig: size mass relation} illustrates the star cluster population 20 Myr after starburst (a) at $z=9.9$ (top row) and the surviving population by $z=8.0$ (bottom row). The left and centre panels show $r_{\rm half\:mass}$ and central surface densities ($\Sigma_0$) as functions of $m_{\rm star \: cluster}$, while the right panels depict the $\Sigma_0 - r_{\rm half\:mass}$ relation. For comparison, high-$z$ star cluster observations -- both photometric \citep{vanzella_jwstnircam_2023, adamo_bound_2024} and spectroscopic \citep{mowla_formation_2024} -- are shown in the top row (see respective papers for similar plots). Here, $\Sigma_0$ is derived by fitting the projected density profile, $\Sigma(r) = \Sigma_{\mathrm{bg}} + \frac{\Sigma_0}{1 + (r / r_{\mathrm{core}})^{\alpha}}$, yielding the background surface density $\Sigma_{\rm bg}$, core radius $r_{\rm core}$, and power-law index $\alpha$ \citep{king_structure_1962}. Errors in $\Sigma_0$ arise from the non-linear least-squares fit.

\begin{figure*}
    \includegraphics[width=\textwidth]{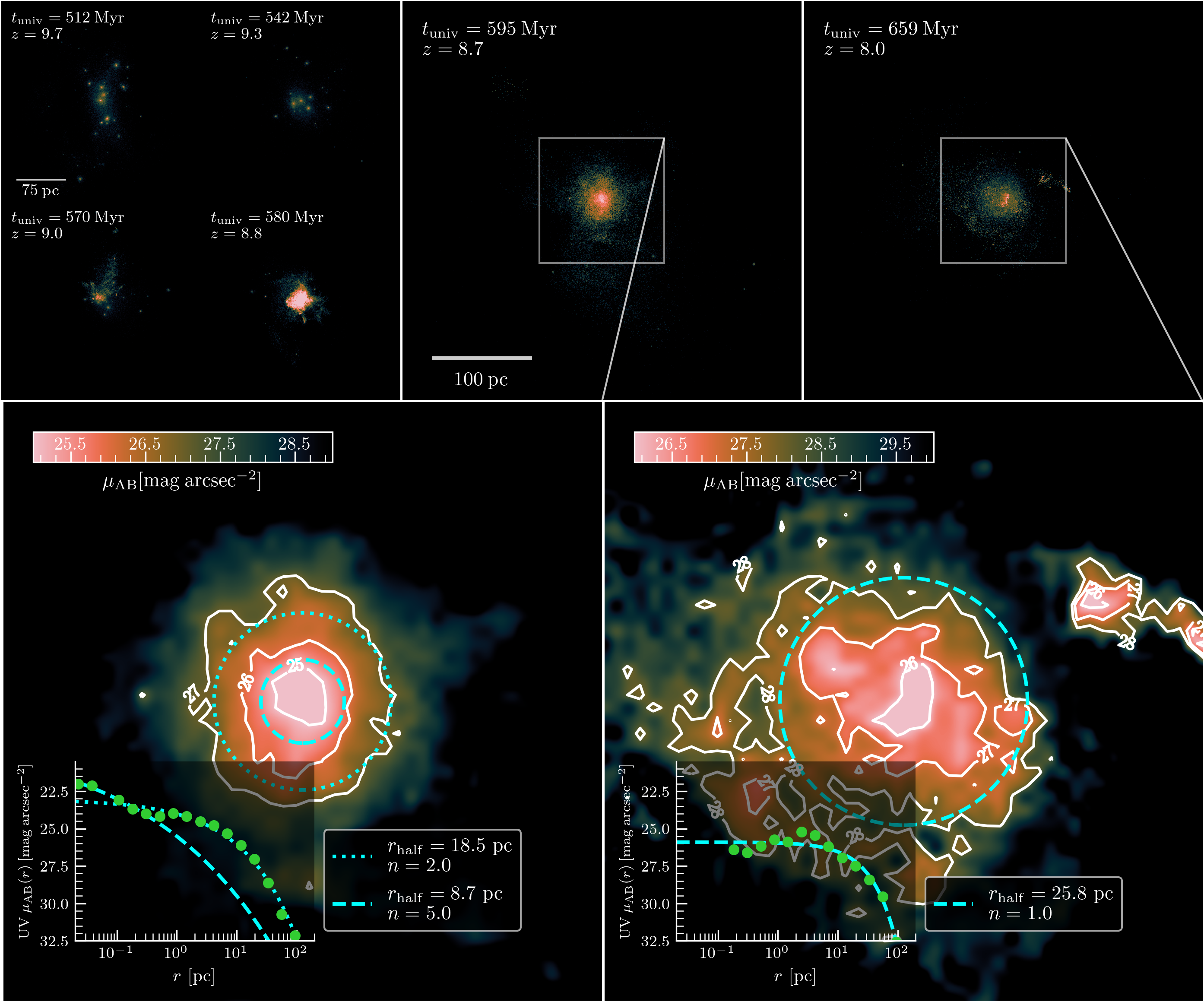} 
    \caption{A time-lapse of the VSFE galaxy leading up to (top left), during (top centre), and after (top right) the formation of the central NSC, coloured according to the redshift-corrected stellar surface brightness (absolute AB magnitudes at 1500 $\text{\AA}$ rest-frame). We see the galaxy transition from a star cluster-dominated galaxy to a primarily bulge-dominated one during $t_{\rm univ}$ = 580 - 595 Myr. We then show a zoomed inset view of the central ($\sim 100$ pc) region of the galaxy right after the bulge was formed from starburst (b) and its state near the end of the simulation (bottom left and right, respectively). For reference, we show surface brightness contours in these insets along with the half-light radii extracted from fitting the surface brightness of the entire galaxy with Sersic profiles (Eq. \ref{eq: sersic surface brithness}, see inset of these panels.)}   
    \label{fig: nsc seeding} 
\end{figure*}

After starburst (a), the galaxy predominantly comprises of bound (see colour bar), compact ($r_{\rm half-mass} \lesssim 3$ pc), and dense ($\Sigma_0 \sim 10^{2 - 4} ~ {\rm M_\odot \: pc^{-2}}$) star clusters. However, by the end of the simulation ($t_{\rm univ} = 659 ~{\rm Myr}$, $z \sim 8.0$), fewer bound star clusters survive and those that do tend to have masses no more than $m_{\rm star-cluster} \sim 10^4~{\rm M_\odot}$ and $r_{\rm half-mass} \lesssim 2$ pc (second row of Figure \ref{fig: size mass relation}). The NSC is an outlier, having $r_{\rm half\: mass}\sim 10$ pc and amassing nearly a few $10^5~{\rm M_\odot}$. For comparison, the bottom row of Figure \ref{fig: size mass relation} also includes properties of nearby ($z \sim 0$) NSCs from \cite{neumayer_nuclear_2020} as dots in the mass-radius, mass-density, and radius-density planes, along with extrapolated fits (note,  $\Sigma_0$ has a weak dependence on $r_{\rm half-mass}$). While the NSC's radius is roughly twice and $\Sigma_0$ about a quarter of the expected values, it still lies within 0.6 dex of the relationship fitted to NSCs. Furthermore, these trends were fitted for a sample of NSCs at $z \sim 0$, while the properties of the NSC and star clusters we show in Figure \ref{fig: size mass relation} are at $z = 8.0$. 

The existence of an NSC significantly impacts the surviving star cluster population. By $z = 8.0$, the grey circles in Figure \ref{fig: size mass relation} highlight a larger population of unbound clusters compared to the top row. These objects have larger sizes ($r_{\rm half\:mass} \gtrsim 1$ pc), lower masses ($m_{\rm star\:cluster} \lesssim 10^4 ~{\rm M_\odot}$), and lower central densities ($\Sigma_0 \lesssim 10^3 ~{\rm M_\odot \: pc^{-2}}$), indicative of dynamical perturbations \citep{Kruijssen2012}. Most bound clusters are remnants of starburst (a), as the NSC’s formation creates a tidally crowded environment that disrupts clusters at birth. Additionally, the formation of a centrally dominant object imposes dynamical friction causing star clusters with wider orbits to gradually migrate to the galactic centre, further contributing to the dissolution of the star clusters and further mass growth of the NSC \citep[\eg~ see][]{Gao2024}.

The VSFE model captures diverse star cluster populations, including those (i) nearly $10 \times$ more massive than the largest clusters in the HSFE and LSFE models of \citetalias{garcia_star_2023}, formed at higher SFEs ($\sim 80 ~{\rm\%}$), and (ii) those formed in lower-mass, metal-enriched clouds with SFEs $< 10$\%. As star formation in the galaxy transitions from (i) to (ii), we observe the seeding and growth of an NSC.

\subsection{Seeding a nuclear star cluster} \label{ssec: nsc seeding}

This section examines the formation of the NSC in detail. Figure \ref{fig: nsc seeding} presents the observed rest-frame UV surface brightness ($\mu_{\rm AB}$) of the galaxy. The top left panel shows 200 pc stamps of the star-forming region after starburst (a) when star clusters dominated the galaxy's light and its $\sim$80 Myr evolution leading to starburst (b), which seeds the NSC (top centre). By $z = 8.0$, the galaxy morphology (top right) becomes UV-dominated at the centre. The bottom row of Figure \ref{fig: nsc seeding} provides a detailed view of the central regions in the top centre and right panels, with surface brightness contours overlaid.

\begin{figure*}
    \includegraphics[width=\textwidth]{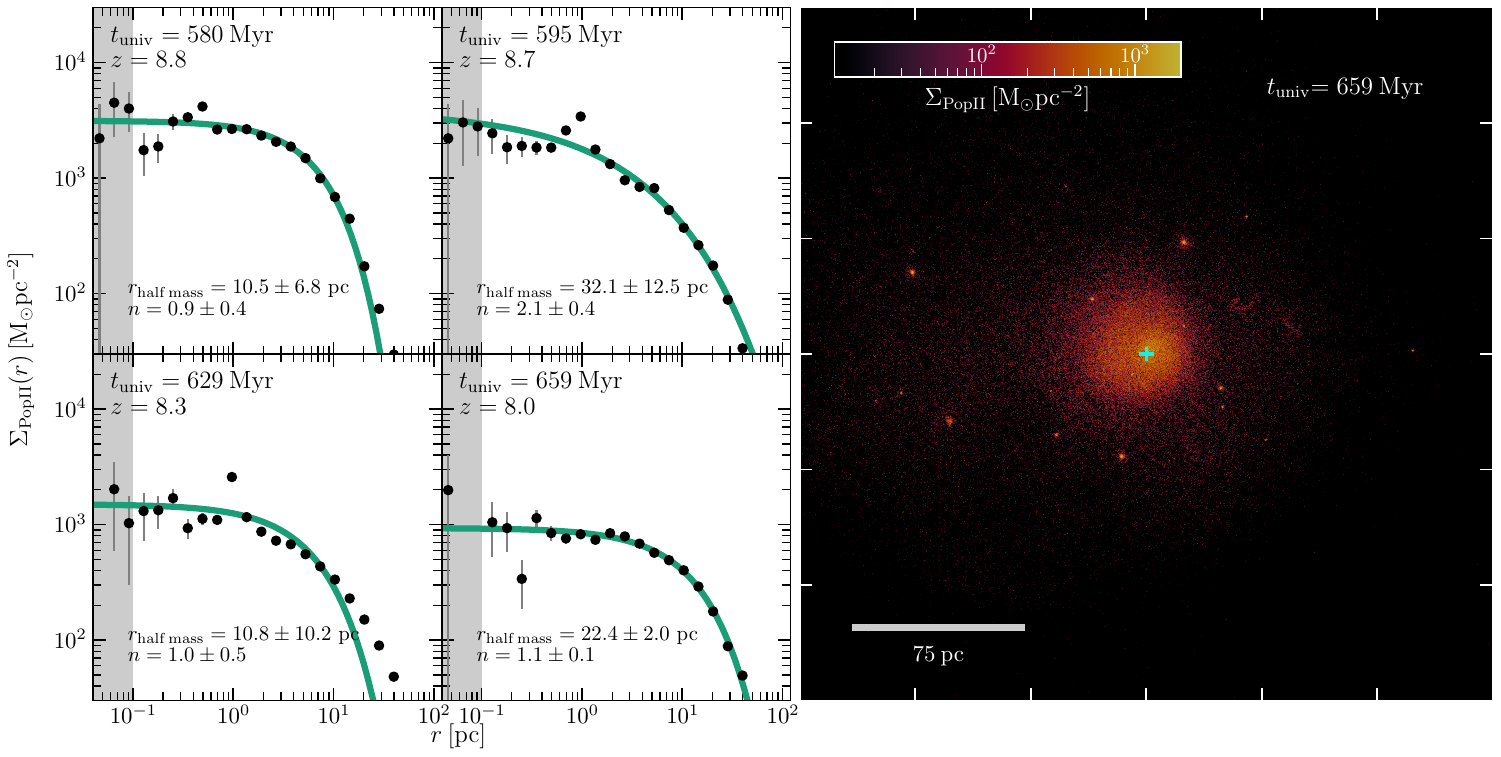}
    \caption{Evolution of the galaxy stellar surface density profiles $\Sigma_{\rm Pop II} (r)$ after starburst (b) (left panels) fitted with a Sersic density profile (see text for details) centred on the NSC which is marked by a cyan crosshair in the projected stellar densities $\Sigma_{\rm Pop II}$ shown on the right panel. The right panel corresponds to the bottom right surface density profile taken at $z = 8.0$. The errors depict Poisson noise.} 
    \label{fig: nsc profiles} 
\end{figure*}

Identifying NSCs is typically ambiguous both observationally and theoretically, especially near the edge cases of the size and mass distributions of NSCs \citep{neumayer_nuclear_2020}. While the properties of the NSC formed will certainly evolve further (recall, the galaxy in this work is expected to grow into $10^{10}~ {\rm M_\odot}$ by $z = 0$), this is where the NSC lies at $z=8.0$: it is near the low-mass end of the NSC mass-to-galaxy-mass (Figure \ref{fig: nsc mass gal}), radius-to-mass (Figure \ref{fig: size mass relation}, bottom left), and density-to-mass (Figure \ref{fig: size mass relation}, bottom right) relations. Nonetheless, a canonical observational marker is the steeping of surface brightness profiles near the central region of a galaxy \citep[e.g,][]{lambert_systematically_2024}. Accordingly, we fit the surface brightness of the entire galaxy after the formation of the NSC with a Sersic profile \citep{sersic_influence_1963}: 
\begin{equation}
\mu_{\rm AB}(r) = \mu_e + \dfrac{2.5 b_n}{\ln(10)}\left[\left(\dfrac{r}{r_{\rm half}}\right)^{1/n} - 1\right],
\label{eq: sersic surface brithness}
\end{equation}
where $r$ is the radial distance from the centre of the galaxy, $\mu_{e}$ is the central brightness, $n$ is the Sersic index, $b_{\rm n} = 1.9992n - 0.3721$ \citep{Capaccioli1989}, and $r_{\rm half}$ is the characteristic radius encompassing half of the galaxy's total (in this case, UV) luminosity \citep{caon_shape_1993}. Right after the NSC was formed, we see a slight steepening in the surface brightness near ($\lesssim 3$ pc) the galaxy's centre. We quantify this in Figure \ref{fig: nsc seeding} (bottom left) with a double-Sersic profile, which fits the NSC (with $r_{\rm half} = 8.7$ pc and $n=5.0$) and the host galaxy ($r_{\rm half} = 18.5$ pc and $n=2.0$) components separately. 

Dashed and dotted circles in Figure \ref{fig: nsc seeding} represent the characteristic radii of the NSC and galaxy components, respectively. The starburst drives the slight steepening in $\mu_{\rm AB}$ ($\sim 2.5$ dex) as the bright, young stellar population and higher central densities amplify UV emission near the centre. UV surface brightness is highly variable due to stochastic star formation, causing the galaxy to dim significantly over $\sim10$ Myr (Figure \ref{fig: nsc seeding}, bottom right), making the NSC less distinguishable in radial UV profiles. By $z = 8.0$, stars formed offset the centre after starburst (b) dominate UV emission, slightly increasing the fitted half-light radius to $r_{\rm half} = 25.8 ~{\rm pc}$ (Figure \ref{fig: nsc seeding}, bottom right).   

Given the variability in the galaxy's UV surface brightness, we also use a Sersic density profile: $\Sigma(r)=\Sigma_0\exp$ $\left[-b_n (r/r_{\rm half\:mass})^{1/n}\right]$ \citep{vitral_precise_2020}. Figure \ref{fig: nsc profiles} (left panels) shows the evolution of the stellar surface density profiles for the entire galaxy, $\Sigma_{\rm Pop II}(r)$, after starburst (b) along with a non-linear least-squares fit. The galaxy has a fluctuating half-mass radius $r_{\rm half \: mass} \sim 20 - 30$ pc, with a stable Sersic index of $n \sim 1 - 2$. Slight over-densities ($\sim 3\times$ higher than the fit) near the centre ($r \lesssim 1$ pc) are attributed to compact star clusters in-spiralling toward the NSC. This is seen qualitatively in the stellar surface densities depicted in the right panel. At $z = 8.8$, right after the NSC was seeded, the central density reaches as much as a few $10^3 ~{\rm M_\odot \: pc^{-2}}$. However, we see that the densities decrease slightly at $z =8.0$, suggesting that the relatively low mass of the NSC and the dynamical heating of the broader galactic environment of the cluster prevent its core collapse (and subsequent formation of an inner cusp in the surface density profiles) \citep{merritt_evolution_2009}. Note, however, that this may also be a resolution effect caused by the 0.15 pc gravitational softening in our simulations which may artificially prevent core collapse.

\begin{figure*}
    \includegraphics[width=\textwidth]{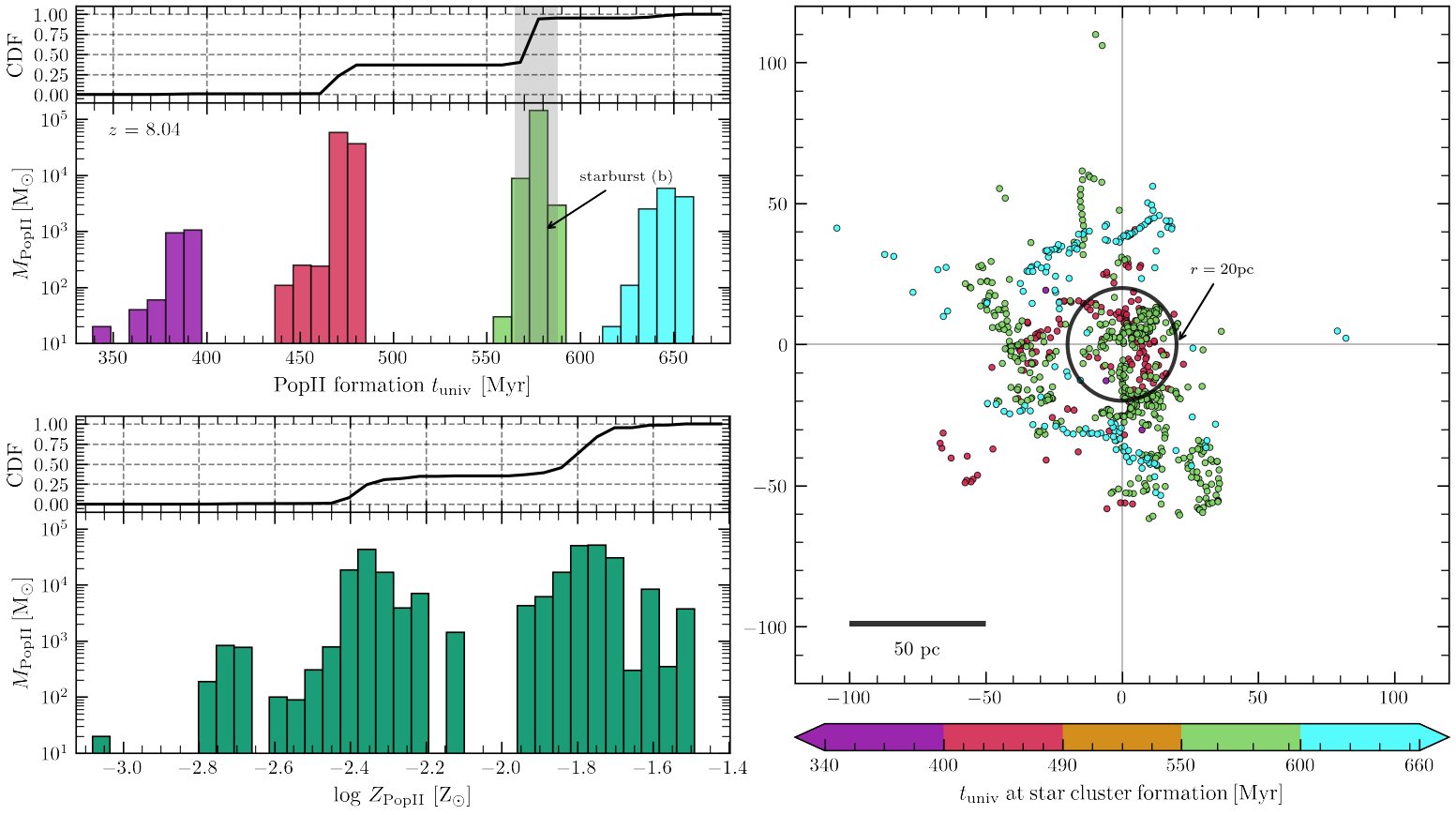}
    \caption{Stellar population that comprises the NSC. We show a mass distribution of Pop II stars ($M_{\rm Pop II}$) comprising the NSC by the end of the simulation  (top left), highlighting starburst (b) when the majority of the NSC mass was formed. Similarly, we also show the distribution of Pop II stellar mass per given metallicity bin in the NSC (bottom left). The right panel shows the positions of clouds identified in the simulation at the time they reach the critical density for star formation. We show their physical positions with respect to the centre of the galaxy (see text for the definition) and colour them based on their birth epoch (clouds that form stars during starburst (b) are coloured green). To guide the reader's eye, we indicate a circle with a radius of 20 pc which is $\sim 2 \times$ the typical half-mass radius of the NSC (refer back to Figure \ref{fig: nsc profiles}).}   
    \label{fig: nsc pop} 
\end{figure*}

\begin{figure}
    \includegraphics[width=\columnwidth]{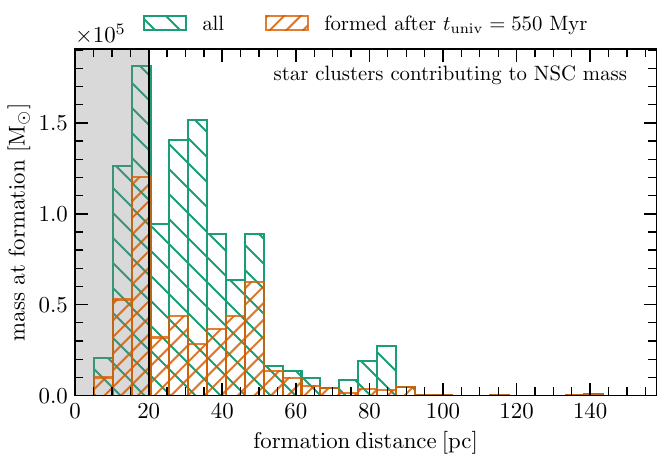}
    \caption{Distribution of formation distance from the centre of the galaxy/NSC for all star clusters (green) that have contributed at least a fraction of their stellar mass to the NSC. We show a subset of this distribution (orange) only for those formed at $t_{\rm univ} > 550 ~{\rm Myr}$, during and after starburst (b). In situ NSC mass contribution is depicted by the orange distribution within the grey-shaded region, indicating star formation within $< 20 ~{\rm pc}$ of the centre.}   
    \label{fig: nsc progenitors} 
\end{figure}

\subsubsection{Stellar populations in nuclear star clusters} \label{sssec: Stellar populations in nuclear star clusters}

We now examine the formation mechanism of the NSC by analyzing its stellar populations. Figure \ref{fig: nsc pop} (top left) shows that the NSC contains stars from all star-forming epochs, including $\sim$ 300 Myr-old stars predating the first starburst at $t_{\rm univ} \sim 360$ Myr. Looking at the cumulative distribution function (CDF), the NSC is partly made (around 40\%) by stars formed in predominantly bound star clusters (masses ranging from a few 100 to $10^5 {\rm M_\odot}$) during starburst (a) at $z \sim 10$. These first-generation star clusters underwent some tidal disruption and dissolution during a $\sim 80 ~{\rm Myr}$ period of quiescence. A second starburst followed, with lower SFEs ($\lesssim$35\%, see Figure \ref{eq: SFE}), producing less bound clusters \citep{kroupa_formation_2001, pfalzner_expansion_2013, Banerjee2018ASSL..424..143B}. Afterwards, star formation proceeds in metal-enriched, highly-fragmented clouds, collectively contributing over 50\% of the NSC mass by $z = 8.0$. Hence, the NSC grows most of its mass through the accretion of the less massive, unbound star cluster population formed in burst (b) as opposed to mass contribution from the evaporation of massive star clusters formed during starburst (a).

We see a similarly multi-peaked distribution looking at the stellar metallicities (Figure \ref{fig: nsc pop}, bottom left), which contain star particles with metallicities as low as $Z_{\rm Pop II} = 8 \times 10^{-4}~{\rm Z_\odot}$ and recently-formed stars of metallicity as high as $Z_{\rm Pop II} = 3 \times 10^{-2}~{\rm Z_\odot}$. Moreover, we mention the minor detail that there is not a direct correspondence between the $t_{\rm univ}$ and $Z_{\rm Pop II}$ distribution, a point we made earlier when looking at Figure \ref{fig: cloud metals}. The NSC consists of stars from all star-forming periods; however, the CDF in the bottom left panel of Figure \ref{fig: nsc pop} shows that it is predominantly ($> 50$\%) made of stars with $Z_{\rm Pop II} > 10^{-2}~{\rm Z_\odot}$. 

The oldest constituent stars in the NSC are a couple of star particles from the first Pop II star cluster formed at $z = 12.95$ ($t_{\rm univ} \sim 350 ~{\rm Myr}$) with metallicities $Z_{\rm Pop II} \sim 2 \times 10^{-4}~{\rm Z_\odot}$. This large metallicity spread ($ -3.5 < \log (Z / {\rm Z_\odot}) < -1.5$) of the NSC, when compared to those within individual star clusters shown in Figure \ref{fig: cluster zdist}, is further confirmation that the NSC is formed -- for a large faction of its mass -- by accreting older star clusters.

\subsubsection{In situ star formation versus star cluster in-spiral} \label{sssec: In situ star formation}

We examine the origin of the NSC further by looking at the sites of formation for all stars in the NSC, depicted in the right panel of Figure \ref{fig: nsc pop}. This figure shows the physical positions of star-forming clouds w.r.t. the centre of the galaxy at the time the cloud reaches $n_{\rm H, crit}$. Immediately, we see that all clouds that contributed mass to the NSC formed stars within 200 pc (physical) of the galaxy's centre. Since the virial radius ($r_{\rm vir}$) of this halo grows from around $r_{\rm vir} \sim 1.2 - 3.4$ kpc throughout the simulation, star formation occurs exclusively within $0.1r_{\rm vir}$ from the halo centre. Furthermore, the spatial distribution of the sites of cloud formation shows that star formation, especially at later times, is clustered in elongated structures resembling filaments or shells.

The circle in the right panel of Figure \ref{fig: nsc pop} encloses a central region with a radius of 20 pc, roughly two times the $r_{\rm half\:mass}$ of the NSC at $z = 8.0$ (Figure \ref{fig: size mass relation}). If we define star formation taking place within this region right before and after starburst (b) ($t_{\rm univ} > 550 ~{\rm Myr}$) to be in-situ star formation, we conclude that while there is some in situ star formation that contributed to the growth of the NSC, the majority of the star clusters were formed ex-situ. To confirm this, Figure~\ref{fig: nsc progenitors}  shows the distribution of (3D) physical distances between the galaxy's centre and the clouds' position at the time of formation. This Figure shows the distances of the newly-formed star clusters from the centre, weighted by that cluster's mass at formation. That is, not all stars from a star cluster necessarily become members of the NSC. Only $\sim 20$\% total by mass of the green distribution shown in Figure \ref{fig: nsc progenitors} have made it to the NSC. This fraction should sound familiar given that nearly all star-forming periods contributed to the growth of the NSC, which comprises around 20\% of the galaxy's total mass by $z = 8$ (see Section \ref{sssec: Seeding of a nuclear star cluster}). Regardless, a small fraction by mass -- around 17\% -- of the NSC donor star clusters formed stars in situ (after 550 Myr and within 20 pc of the galaxy's centre). This in situ fraction is shown as the shaded orange distribution in Figure~\ref{fig: nsc progenitors}.

In addition to the small NSC mass contribution from in situ star formation by $z=8.0$, we also note that star formation post starburst (b) occurs almost exclusively just outside of the NSC (cyan points in Figure \ref{fig: nsc pop}, right panel). This is likely due to feedback from the preceding starburst preventing the gas from reaching $n_{\rm H, crit}$ near the centre of the galaxy. The star-forming clouds are distributed in a ring-like structure that resembles a nuclear stellar ring \citep{comeron_ainur_2010, brandl_high_2012, ma_connections_2018}. However, this structure is probably unstable dynamically, and a stellar ring will likely be short-lived.

The results above suggest that the formation of the NSC comes from two pathways: (i) old star clusters that eventually evaporated due to dynamical relaxation and/or disrupted due to external perturbations as they migrate inward to the centre of the galaxy and (ii) open star clusters that formed in-situ (within 20 pc) near the centre of the galaxy. At this stage of the galaxy's evolution, the dominant pathway for NSC growth is via star cluster accretion and dissolution while in-situ star formation marginally contributes to the mass at merely 17\%.  

\section{Discussion}\label{sec:Discussion}

To our knowledge, this is the first cosmological RHD simulation of a galaxy that forms both massive bound star clusters and a central object resembling an NSC at $z \gtrsim 8.0$. We discuss our results within the broader effort to study high-redshift bound star clusters, now being observed by JWST, and postulate potential links to NSC formation.

\subsection{Comparison to recent observations of magnified star clusters at high redshift}

Here, we draw comparisons between the star clusters formed in our simulations and those recently observed by the JWST aided by strong gravitational lensing. We refer to Figure \ref{fig: size mass relation} for relevant star cluster properties and scaling relationships in our simulation. 

\cite{vanzella_jwstnircam_2023} presented NIRCam photometric observations of Sunrise Arc, a lensed galaxy at $z_{\rm phot} = 6.0 \pm 0.2$. The entire galaxy has an estimated stellar mass of $M_\star \sim 0.3 - 2.0 \times 10^9 ~{\rm M_\odot}$ with roughly 10 to 30\% of its mass locked in 6 detected star clusters with individual masses $m_{\rm star \: cluster} \sim 10^{6-7} ~ {\rm M_\odot}$. These star clusters are young (ages $\lesssim 5 - 30 ~{\rm Myr}$), compact (effective half-light radii $\sim 1 - 25 ~{\rm pc}$), and have high stellar surface densities: $\Sigma_\star \sim 1.5-12.0 \times 10^{3} ~{\rm M_\odot \: pc^{-2}}$, with one cluster in particular reaching up to a few $10^6 ~{\rm M_\odot \: pc^{-2}}$. 

Another set of observations is of the Cosmic Gems Arc at an even higher redshift ($z_{\rm phot}\sim10.2$) by \cite{adamo_bound_2024}, revealing an even more extreme population of star clusters that -- while having similar masses with 5 identified young star clusters (ages $\sim 9-36$ Myr, $m_{\rm star\: cluster}\sim10^6 ~{\rm M_\odot}$) hosting 30\% of the galaxy's total mass of $M_\star\sim2.4-5.6\times10^7 ~{\rm M_\odot}$ -- have much lower estimated radii ($\lesssim 1 ~{\rm pc}$) and generally higher surface densities $\Sigma_\star > 10^5 ~{\rm M_\odot \: pc^{-2}}$.

Most recently, \cite{mowla_formation_2024} reported spectroscopic observations of the Firefly Sparkler galaxy \citep{mowla_sparkler_2022}. Confirmed at $z_{\rm spec} = 8.296 \pm 0.001$, this galaxy has a stellar mass estimate of $M_\star \sim 5.0 \times 10^{6} - 1.0 \times 10^{8}~{\rm M_\odot}$ and is made of up to 57\% by mass from compact ($\lesssim 4-7$  pc), young (ages $\sim 2 - 8$ Myr assuming instantaneous SSP burst) star clusters with masses $\sim 10^{5-6} ~{\rm M_\odot}$ and densities $\Sigma_\star \gtrsim 10^{3 -4}~{\rm M_\odot\: pc^{-2}}$. 

The relatively young star cluster population formed in our simulation (ages $\lesssim 20$ Myr at $z = 9.9$) bear most resemblance to those presented in \cite{mowla_formation_2024}, though they tend to be less massive ($m_{\rm star \: cluster} \lesssim 10^5 ~{\rm M_\odot}$, more compact ($r_{\rm half\: mass} \lesssim 3$ pc), and less dense ($\Sigma_0 \lesssim 2 \times 10^4 ~ {\rm M_\odot\: pc^{-2}}$) than the current observations. However, because the sources in \cite{mowla_formation_2024} are unresolved, they are likely to be smaller and denser. We also highlight the fact that although our simulations have a sub-pc resolution ($\Delta x_{\rm min} = 0.15$ pc at $z=9$), higher resolution is likely required to not only resolve smaller star clusters but also to reliably track the dynamics and maintain the high densities of these parsec-sized objects. 

\begin{figure}
    \includegraphics[width=\columnwidth]{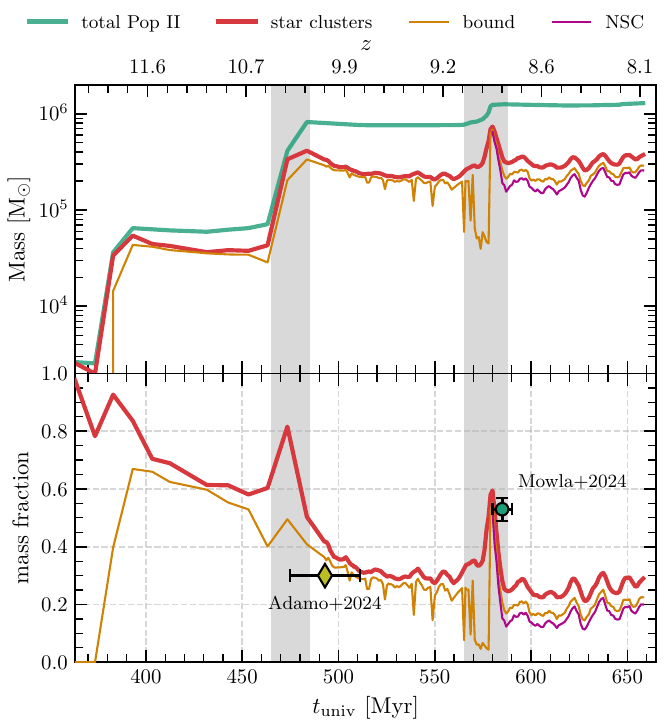}
    \caption{Star cluster formation efficiencies. In the top row, we show the total mass of Pop II stars similar to Figure \ref{fig: SF History} (note, the grey regions are also adopted from the same figure) in addition to the total mass residing in star clusters (both open and bound), bound star clusters, and the NSC after it was initially seeded during starburst (b). In the bottom row, we calculate the mass fraction of these quantities with the overall Pop II stellar mass of the galaxy (green line above), showing them with cluster formation efficiencies reported by \citep{adamo_bound_2024} and \citep{mowla_formation_2024}. Note that, for the observations (symbols), the depicted times have been slightly changed with respect to the observed redshifts such that the observed ages of the star clusters are correctly reproduced assuming they form during the the nearest starburst in our simulated galaxy.}
    \label{fig: bound fractions} 
\end{figure}

The galaxy in this work reaches a maximum bound star cluster formation efficiency (CFE), which is the instantaneous fraction of stars observed in a star-forming galaxy formed in star clusters, of more than 60\% during the formation of the first few star clusters at $z\sim 11.6$. The CFE decreases to nearly 50\% during the first starburst at $z\sim10$ (around 75 Myr after) when the galaxy had a mass of $M_\star = 8.4 \times 10^5 ~{\rm M_\odot}$. Figure~\ref{fig: bound fractions} shows these mass fractions in addition to the mass in the NSC that forms near the end of the simulation. By redshift $z=8.29$, the galaxy's total mass is $M_\star = 1.4 \times 10^6 ~{\rm M_\odot}$ with a CFE of 25\%. The CFE remains relatively stable at 20\% near the end of the simulation at $z \sim 8.0$, with the NSC dominating much of the total mass of the bound systems in the galaxy, roughly 75~Myr after starburst (b) and 175 Myr after starburst (a). The variability in the measured CFE here is consistent with observations, which suggests CFEs ranging between 40 - 60 \% \citep{vanzella_jwstnircam_2023, adamo_bound_2024, mowla_formation_2024}. The symbols in Figure~\ref{fig: bound fractions} (bottom) compare our results with CFEs from JWST observations cent \citep{adamo_bound_2024, mowla_formation_2024}. Given that we only have one realization of a simulated galaxy, instead of placing the data points from the observations at the galaxies' observed redshifts, we place them after the nearest starburst: burst (a) or (b). More specifically, we placed them at times such that the horizontal error bars have widths that encapsulate the range of star cluster ages (stated above), with the lower bounds corresponding to the peak of the nearest starburst.  Given this caveat, the CFE predicted by the simulation is within the observed range and is consistent with both observed values.

It is important to note that the star clusters observed at $z>6$ by JWST are hosted in galaxies with halo masses greater than in our simulation, typically Milky Way progenitors or more massive galaxies \citep{mowla_formation_2024}. Therefore, the maximum star cluster mass and the number of GC progenitors in our simulated galaxy is more typical of a lower-mass halo, consistent with the GC-to-halo mass ratios inferred from observations \citep{forbes_extending_2018, eadie_clearing_2022, jones_gas-rich_2023} and theoretical predictions from cosmological hydrodynamic simulations \citep{doppel_modelling_2022} suggesting that lower mass galaxies $M_\star < 10^9 ~{\rm M_\odot}$ have lower GC occupation fractions. 

Some observations \citep[e.g., the Sunburst Arc][]{2024A&A...690A.269R, 2025MNRAS.540.2991A} also indicate higher inferred gas metallicities in star forming clumps (from nebular emission lines) than the ones found in this work. This is also consistent with our galaxy being relatively low-mass and low metallicity with respect to the observed sample of high redshift galaxies \cite[e.g.,][]{mowla_formation_2024}. 
Since in our simulations the formation of dense and strongly bound star clusters is suppressed at later times as the ISM becomes more metal-rich, it's worthwhile speculating on the effect of mass and metallicity on the star cluster population. 

In a more massive galaxy, assuming simple scaling arguments, we expect stronger bursts of star formation and more massive gas clouds for a given power-law slope of the star cluster initial mass function. These scaling relations may balance out the effect of increased gas metallicities, but more simulations of higher mass galaxies need to be run to confirm this hypothesis.

In addition, there are possible pathways for bound star cluster formation in relatively metal-rich galaxies which our simulations may miss. The physical reason for producing compact star clusters, as discussed in \cite{sugimura_violent_2024}, is the high Jeans mass (i.e., hot gas) in the ISM of the first galaxies produced by the suppression of $H_2$ cooling that follows from a strong FUV radiation in low-metallicity gas. A similar effect could be obtained by keeping the collapsing gas clouds hot by heating the gas instead of only suppressing cooling. Possible heating sources that we are missing in our simulations include the effect of X-rays from binary stars, accreting IMBHs, or micro-AGN. Some work along this line is currently underway and will be presented in a future paper.
\\
\subsection{Nuclear star cluster formation}

The existence of a central NSC is a fairly common attribute in most galaxies. The nucleation fraction increases in more massive galaxies ($> 90$\% for galaxies with $M_\star \sim 10^9~ {\rm M_\odot}$) \citep[\eg][]{Sanchez-Jansse2019}, and decreases in low mass galaxies $\sim 10^6 \: {\rm M_\odot}$ at $z\sim 0$ \citep{carlsten_wide-field_2020, neumayer_nuclear_2020}. 
Studies as early as those by \cite{tremaine_formation_1975} have proposed in-spiral of GCs due to dynamical friction as a natural formation mechanism for NSCs, with the most massive star clusters especially vulnerable to such fate \citep{neumayer_nuclear_2020}. This scenario is well established, in part due to observations of metal-poor stars at the centre of dwarf galaxy NSCs \citep[\eg][]{Alfaro-Cuello2020, Fahrion2020A&A} that were proposed likely to be tidal remnants of these in-spiralling objects \citep{perets_age_2014}. Indeed, this scenario is likely the most common pathway for NSC growth for low mass ($\lesssim 10^9 ~{\rm M_\odot}$) galaxies presuming they contain a substantial population of star clusters with masses $\gtrsim 10^5 ~{\rm M_\odot}$ that can grow the NSC mass \citep{agarwal_nuclear_2011, neumayer_two-dimensional_2011, neumayer_nuclear_2020}. However, the amount of mass contributed to the NSC via this process ultimately depends on the high-mass truncation of the ICMF, which in our case is roughly $m_{\rm CMF,\: trunc} \sim 10^5~{\rm M_\odot}$, with higher masses leading to more massive and extended NSCs that more comfortably fall within scaling relations \citep{agarwal_nuclear_2011}.   

Otherwise, in-situ star formation fueled by the infall of dense gas \citep{loose_bursts_1982, mihos_dense_1994} near the centre of the galaxy is needed to reproduce kinematics and observed luminosity functions of NSC-hosting galaxies \citep{hartmann_constraining_2011, antonini_dissipationless_2012}. Predictions by semi-analytical models and simulations suggest that it is rather a combination of the two \citep{gnedin_co-evolution_2014, brown_nuclear_2018}. For example, a \textit{wet migration} scenario (as opposed to ``dry'' mergers and dissolution of GCs) in isolated dwarf galaxies presented in  \cite{guillard_new_2016} posits that a YMC can form stars in a gas-rich disk and maintain a gas reservoir for star formation as it in-spirals to the nucleus of the galaxy. Most recently, a starburst-induced in-situ formation of an NSC has been explored in dwarf galaxies at $z \sim 1.4 - 0$ by \cite{Gray2024}. Furthermore, a hybrid NSC formation scenario is identified by \cite{vandonkelaar_stellar_2024} using cosmological hydrodynamic simulations of a MW-mass galaxy at $z > 4$ whereby an NSC can partly form (a maximum of 20\% of its mass) from this mechanism. The rest of the NSC mass is supplied by both disc and bar-funnelled gas into the galactic centre \citep{vandonkelaar_stellar_2024}.  

We see a similar hybrid formation scenario in our simulation, albeit for a dwarf galaxy progenitor occurring at a much earlier time -- within 1 Gyr of cosmic history -- with the majority ($\gtrsim 80$\%) of the mass coming from the infall and disruption of the first star clusters suggesting that this is the dominant mechanism during the initial phase of NSC seeding and growth. This work hints at a possible pathway for the transition between the clustered star formation we see at high-$z$ \citep{vanzella_jwstnircam_2023, adamo_bound_2024, mowla_formation_2024} to the seeding of a centrally dominant NSC at redshifts as early as $z \sim 8.0$, about 700 Myr after the Big Bang.

\section{Conclusions}\label{sec:Conclusions}

We present a high-resolution cosmological RHD zoom-in simulation of a typical mass galaxy evolving during the first 700~Myr of cosmic history (in a $10^{8-9}$~M$_\odot$ DM halo), with UV luminosity comparable to the faintest galaxies observed by JWST at similar redshifts. 
The work presented in this paper is the natural extension of our previous study (\citetalias{garcia_star_2023}), focusing on how changing the star formation efficiencies in molecular clouds affects the formation and evolution of the first star clusters at redshifts $z>8$. These simulations resolve (at $\sim 0.15$~pc resolution) star cluster formation into individual massive ($m_* = 10 ~{\rm M_\odot}$) star particles that emit radiation and inject thermal feedback from CCSNe. 
In \citetalias{garcia_star_2023}, we compared two models: the HSFE model with 70\% efficiency and the LSFE model with 35\% efficiency in star-forming clouds.
In this study, we implement a more physically motivated sub-grid model (the VSFE model) where the SFEs in star-forming clouds vary depending on their densities, masses, and metallicities. \cite{he_simulating_2019} derived the VSFE model we adopted in the present simulations using a grid of high-resolution (AU-scale) RMHD simulations of turbulent molecular clouds. 
Using this multi-scale approach, our study finds the following: 
\begin{enumerate}
    \item Using a star formation model with cloud-dependent SFEs, a dwarf galaxy ($M_h \sim 10^{10} ~{\rm M_\odot}$ at $z = 0$) progenitor produces, by $z\sim 8$, a population of small (0.1 - 3 pc) bound star clusters with masses reaching $\sim 10^{5} ~{\rm M_\odot}$ (nearly $10\times$ more massive than what was initially presented in \citetalias{garcia_star_2023}). These star clusters have stellar surface densities ranging between a few $100 - 2 \times 10^4 ~ {\rm M_\odot \: pc^{-2}}$ and are formed in dense ($\Sigma_{\rm cloud} \sim 2 \times 10^3 ~ {\rm M_\odot \: pc^{-2}}$), metal-poor ($Z_{\rm cloud} \sim 10^{-3} ~{\rm Z_\odot}$) star-forming clouds at redshifts $z = 12.0 - 9.0$. The SFEs in these clouds reach values as high as 80\%. They have similar surface densities to those discovered in JWST observations of strongly lensed galaxies at $z > 6$, but generally are less massive and more compact. Nevertheless our results provide important insights on the physics of dense cluster formation and the results are consistent with observations if we consider simple scaling arguments \citep[e.g.,][]{forbes_extending_2018, eadie_clearing_2022, jones_gas-rich_2023} to adjust for the smaller mass of our simulated host galaxy (a progenitor of  a $10^{10} ~{\rm M_\odot}$ halo at $z = 0$) with respect to the hosts of observed star clusters that typically reside in Milky Way progenitors.
   
    \item Adopting a cloud-dependent SFE model also increases the total stellar mass in the galaxy and the stochasticity of star formation compared to the constant high (70\%) and low (35\%) SFE models. The galaxy with this model experiences the longest period of quiescence (80 Myr, as opposed to HSFE's 70 Myr), and the highest peaks of SFR, doubling the HSFE model's 0.12 ${\rm M_\odot \:yr^{-1}}$. This effect is non-linear and feedback-mediated, enhancing larger starbursts and suppressing smaller star-forming episodes.
    \item Cloud's SFEs decrease over time due to higher metal enrichment which allows clouds to fragment and form less massive (a few 100 to $10^3~{\rm M_\odot}$) unbound/open star clusters, causing the star cluster mass function -- well described by a power law: d$N/ {\rm d\log} \: m_{\rm star\: cluster} \propto m_{\rm star\: cluster}^\Gamma$ -- to gradually steepen from $\Gamma = -0.3$, after the first starburst episode, to $\Gamma = -1.1$ after the formation of a central nuclear star cluster. The slope $\Gamma =-0.3$ is significantly flatter than typically observed at lower-redshifts ($\Gamma \sim -1$) but in good agreement with a recent analysis of the star clusters in the "Cosmic Gems" arc at $z=10$ \citep{Vanzella2025}. The galaxy-scale bound cluster formation efficiency decreases over time with a maximum of 60\% at $z = 12.0$ to 20\% at $z = 8.0$.
    \item We also find that each star cluster has a metallicity spread of 0.05 - 0.1 dex, roughly scaling with the cluster mass, due to inhomogeneities or gradients in the gas metallicity of the natal environment (i.e., pre-enrichment inhomogeneities).
    \item The star cluster system at $z = 9.69$ is produced by the rapid fragmentation of infalling gas filaments that form stars before reaching the halo centre. Hence, the clusters orbit within 150~pc of the galaxy centre, which remains largely devoid of stars. However, at redshift $z\sim 8.7$, a nuclear star cluster (NSC) (mass of $\sim 2 \times 10^5 ~{\rm M_\odot}$ and half mass radius of 10 pc corresponding to a central density of $10^3 ~ {\rm M_\odot \; pc^{-2}}$) forms from the in-spiral and dynamical disruption of star clusters, with the majority of the total NSC mass (83\%) coming from unbound clusters formed at lower ($\lesssim 30$\%) SFEs. The formation of the central star cluster also influences the population of bound star clusters formed at high SFEs, causing them to migrate inwards and lose mass. A sub-dominant fraction of the NSC mass ($\sim17$\%) comes from in-situ star formation. Forming a compact NSC at such early cosmic times has interesting implications for SMBH seeding models and for interpreting the ``little red dots" population recently discovered by JWST at $z \gtrsim 5$.
\end{enumerate}

The results presented here come with several caveats and are likely to improve with more detailed astrophysical models; \eg ~tracking metal yields from various sources and more thorough consideration of pre-SN feedback like stellar winds \citep{Andersson2024}. Furthermore, higher numerical fidelity is also crucial since even though we resolve scales as small as $0.15 ~{\rm pc}$, current observations suggest that the first star clusters are even more compact \citep{adamo_bound_2024, mowla_formation_2024}. The Pop II stars are also represented rather simplistically, with each star particle being $10 ~{\rm M_\odot}$. Our current targets for improvement include: (i) extending the sample of galaxies to include more massive halos, since everything presented thus far is based on the study of one dwarf galaxy analogue; (ii) a more accurate, star-by-star treatment of Pop II stars with masses sampled from an IMF; and (iii) increasing the fidelity of our Pop III star formation model to be consistently determined by cloud-scale properties \citep[\eg][]{Hirano2015} which is crucial in setting the stage for the metallicity evolution of the first Pop II star clusters. We leave these aforementioned improvements for future work.

\section*{Acknowledgments}

We thank the anonymous referee for comments that improved the quality of this paper. FABG acknowledges support from the U.S. Department of Energy, Office of Science, and Office of Advanced Scientific Computing Research, under Award Number DE- SC0025528. This research is also supported by Grants-in-Aid for Scientific Research (KS: 22KK0043, 24H00002) from the Japan Society for the Promotion of Science. The authors acknowledge the University of Maryland's supercomputing resources (hpcc.umd.edu). 
\newpage
\bibliographystyle{mnras}
\bibliography{main}

\end{document}